\documentclass[twocolumn,showpacs,preprintnumbers]{revtex4}
\usepackage{bm} \usepackage{amssymb} \usepackage{amsmath}
\usepackage{graphicx}
\begin{document}
%%%%%%%%%%%
%%%%%%%%%%%
\title{Shear Viscosity of the Outer Crust of Neutron Stars: Ion Contribution}
%%%
\author{O.L. Caballero$^1$, S. Postnikov$^2$, C.J. Horowitz$^1$, and
M. Prakash$^2$}
%\email{lcaballe@indiana.edu}
\affiliation{$^1$ Department of Physics and Nuclear Theory Center,
             Indiana University, Bloomington, IN 47405}
%\author{S. Postnikov}\email{postnik@phy.ohiou.edu}
\affiliation{$^2$ Department of Physics and Astronomy,
             Ohio University, Athens, OH 45701}

%%%%%%%%%%%
\date{\today}
\begin{abstract}
The shear viscosity of the crust might have a damping effect on the
amplitude of r-modes of rotating neutron stars.  This damping has
implications for the emission of gravitational waves.  We calculate
the contribution to the shear viscosity coming from the ions using
both semi-analytical methods, that consider binary collisions, and
Molecular Dynamics simulations.  We compare these results with the
contribution coming from electrons.  We study how the shear viscosity
depends on density for conditions of interest in neutron star
envelopes and outer crusts.  In the low density limit, we find good
agreement between results of our molecular dynamics simulations and
classical semi-analytic calculations.
\end{abstract}
\smallskip

\maketitle

\section{Introduction}
Neutron stars are good resonators where oscillation modes can be
excited. In particular, the r-modes of rotating neutron stars involve
currents associated with very small density variations. These modes
are unstable at all rates of rotation in a perfect fluid star
\cite{AnderssonGW}. The instability is due to the emission of
gravitational radiation, suggesting the possibility of gravitational
wave detection with Laser Interferometer Gravitational-Wave Observatory (LIGO) \cite{Owen-LIGO}. This instability is
expected to spin down newly born hot neutron stars \cite{LindblomGW}.
However, the observation of colder rapidly rotating neutron stars,
suggests the existence of a damping mechanism of the r-mode
instability. Several works have been done to explain this
mechanism. For example, in Ref. \cite{Bildsten2000} the damping
mechanism is suggested to be the result of a viscous layer at the
interface between the solid crust and the fluid core. Other works
discuss the r-mode dynamics of superfluid neutron stars
\cite{Andersson2001,Lindblom2000} finding that a core filled with
neutron and proton superfluids limits the amplitude growth of the
modes \cite{Lee_r-modes}.  In addition, high multipolarity p-mode
oscillations may impact the pulse shape of some radio pulsars
\cite{pulseshape}.  For p-modes the primary restoring force is the
pressure and the modes may be damped by the shear viscosity of the
neutron star crust \cite{Chugunov-shear}.

In the outer crust, the total shear viscosity has contributions from electrons and ions both of which transport momentum. Previous works have
 calculated the electron contribution to the shear viscosity in the
 Born approximation \cite{Flowers1979} and including non-Born
 corrections \cite{Chugunov-shear}.  Recently Horowitz and Berry
 calculated the electron contribution to the shear viscosity of
 non-spherical nuclear pasta phases \cite{viscos_pasta}.  In this work
 we study the dependency of the shear viscosity with the density. We
 calculate the contribution of the ions to the shear viscosity in the
 neutron stars crust by two different methods: via Molecular Dynamics
 (MD) simulations and calculating momentum transport cross
 sections. The first method follows the Kubo formalism \cite{Kubo},
 and calculates the autocorrelation function of the pressure
 tensor. The second method considers binary collisions and allows one
 to consider both the classical and quantum systems. We focus our
 study to the case in which the ions form a dilute Hydrogen One
 Component Plasma (OCP). The conditions of temperature and density are
 chosen to reproduce those of the envelope and outer crust. However,
 the formalism can be applied to calculate other transport properties,
 such as diffusion coefficients. In particular the numerical procedure
 can be extended to study Multi-Component Plasmas (MCP).

This paper is organized as follows:  in Sec. \ref{sec:formalism} we
describe our ion-ion interaction model, the Kubo formalism, and the
semi-analytical procedure to calculate transport coefficients from
transport cross sections. In Sec. \ref{dilute} we present our
numerical and semi-analytical results for the dependency of the shear
viscosity with density. In Sec. \ref{quantum} we discuss quantum
results vs classical ones. In Sec. \ref{electron-shear} we compare our
results with the contribution coming from the electrons, and finally
in Sec. \ref{conclusions} we conclude. The appendix contains the
description of the method employed to calculate the phase
shifts. 

\section{Formalism}
\label{sec:formalism}
\subsection{Ion-Ion Interaction Model}
\label{sec:IIIM}
Electrons in the crust of a neutron star form a very degenerate
relativistic gas that screens the interaction between ions. The ions
form a plasma in the neutralizing electron background. It is
convenient to characterize the strength of the ion interactions in
terms of the Coulomb coupling parameter $\Gamma$, which is defined as
the ratio of the average potential energy to the average kinetic
energy,
\begin{equation}
\Gamma=\frac{Z^2e^2}{a_rT},
\label{Gamma}
\end{equation}
where $T$ is the temperature of the system, which we report in MeV
$(T[$MeV$]=k_BT)$, $Z$ is the charge of the OCP, and
\begin{equation}
a_r=\left(\frac{3}{4\pi n}\right)^{1/3},
\label{ion-sphere}
\end{equation}
is the ion sphere radius.  Finally, $n$ is the ion density.
The plasma frequency for a OCP is,
\begin{equation}
\omega_p =\sqrt{\frac{4\pi nZ^2e^2}{M}}\,,
\label{plasmaT}
\end{equation}
where $M$ is the mass of one ion. Quantum effects are expected to be
important if $T<<T_p$ with $T_p= \omega_p$ the plasma temperature.
 
 We describe the interaction between ions with a Yukawa potential,

\begin{equation}
   V(i,j)=\frac{Z_i Z_je^2}{r_{ij}}e^{-r_{ij}/\lambda_e},
   \label{potential}
\end{equation}
where $r_{ij}$ is the distance between the \emph{i}th and \emph{j}th
ions, $\lambda_e=\pi^{1/2}/(2ek_F)$ is the electron screening length
with the electron Fermi momentum $k_F=(3\pi^2 n_e)^{1/3}$, the
electron density is $n_e=\langle Z \rangle n$, and $e^2=\alpha$ the
fine structure constant. In the OCP case all $Z_i$ correspond to the
same ion species.

\subsection{Ion-ion Autocorrelation Function of the Pressure tensor}
\label{sec:autoc}
We use the Kubo formalism to calculate transport coefficients
\cite{Kubo}.  Kubo showed that linear transport coefficients $L$,
could be calculated from a knowledge of the equilibrium fluctuations
in the flux $J$ associated with the particular transport coefficient,
\begin{equation}
L=\frac{V}{k_BT}\int_0 ^\infty dt \left \langle J(t)J(0) \right \rangle,
\label{kubo}
\end{equation}
where $V$ is the volume of the system, and $T$ its temperature. The
integrand in Eq. (\ref{kubo}) is called the autocorrelation
function. At zero time the autocorrelation function is the mean square
value of the flux. At long times the flux $J(t)$ at time $t$ is
uncorrelated with its value at $t=0$, $J(0)$ and the autocorrelation
function decays to zero.

In particular, to calculate the shear viscosity $\eta$, we have made
use of the autocorrelation function for the pressure tensor $P$,
\begin{equation}
\eta =\frac{V}{3k_BT} \int _0 ^\infty 
\left \langle \sum_{x<y} P_{xy}(t+t_0)P_{xy}(t_0)\right \rangle dt.
\label{eta}
\end{equation}
 The
average is taken over the three off-diagonal components $(x,y)$,
$(y,z)$, $(z,x)$, and over different initial times $t_0$.

Explicitly, the pressure tensor is given by
\begin{equation}
P_{xy}(t)=\frac {1}{V}\left [\sum_j m_jv_{x_j}(t)v_{y_j}(t)+
\frac{1}{2}\sum_{i\neq j}r_{x_{ij}}(t)F_{y_{ij}}(t)\right ].
\label{pxyz}
\end{equation}
Here $v_{x_{i}}(t)$ is the $x$ component of the velocity of the $i$th
ion at time $t$, $r_{x_{ij}}$ is the distance in the $x$ direction
between the $i$th and $j$th ions, and $F_{y_{ij}}$ the $y$ component
of the force between them.

This microscopic description allows us to calculate the viscosity of a
gas, in which case we interpret the component $P_{ij}$ of the pressure
tensor as the mean increase, per unit time and per unit area across a
plane in the $j$ direction, of the $i$th component of momentum of the
gas \cite{Reif}.

\begin{figure}[htbp]
\begin{center}
\includegraphics[width=2.8in]  {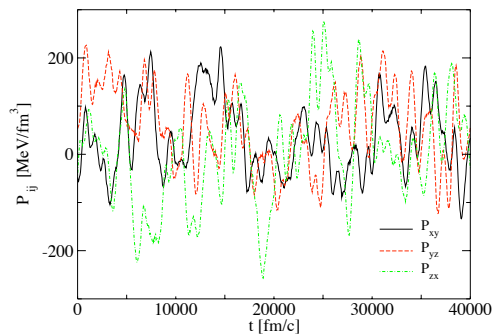}
\caption{(Color online.) The three off-diagonal components of the
pressure tensor Eq. (\ref{pxyz}), corresponding to a simulation of OCP
with $Z=29.4$, $A=88$, a temperature of 1 MeV and an ion density of
7.18 $\times 10^{-5}$ fm $^{-3}$.}
\label{pressureT}
\end{center}
\end{figure}

Figure \ref{pressureT} shows the three different off-diagonal components
of the pressure tensor. We need to calculate the correlation of this
statistical noise. This is the aim of Eq. (\ref{eta}). The components of
the pressure tensor correspond to a simulation of OCP, for $T= 1$ MeV,
$n= 7.18 \times 10^{-5}$ fm$^{-3}$ , $Z= 29.4$, and $A=88$. The total
simulation time is $1.7 \times 10^7$ fm/c and the time step is $\Delta t=
50$ fm/c.

\begin{figure}[htbp]
\begin{center}
\includegraphics[width=3.5in]  {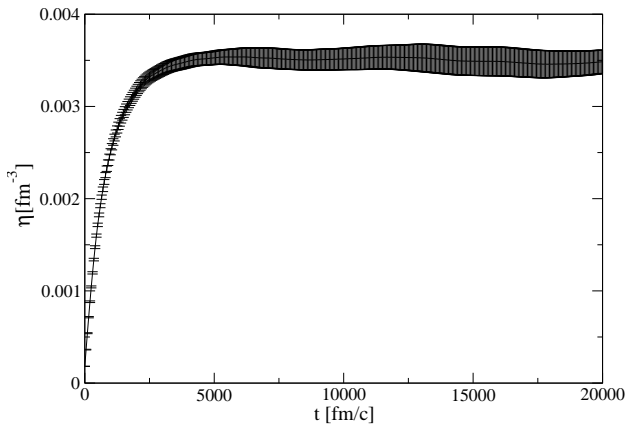}
\caption{(Color online.) Viscosity $\eta$ as a function of the upper
time limit of Eq. (\ref{eta}). In this simulation all ions have
$Z=29.4$, and $A=88$. The temperature is 1 MeV and the ion density is
7.18 $\times 10^{-5}$ fm $^{-3}$.}
\label{vis}
\end{center}
\end{figure}

In principle the integration in Eq. (\ref{eta}) will require an infinite
upper limit. In practice, we integrate up to a finite time
$t_{up}$. The choice of this upper limit requires caution. We want to
find autocorrelations in the pressure tensor for long times, but at
the same time if we let $t_{up}$ be too long the system becomes
uncorrelated, and the statistical errors will increase. To solve this,
we assume that after a perturbation the system relaxes
exponentially. Then we find the relaxation time for the pressure
tensor, and we integrate Eq. (\ref{eta}) for up to few times this
relaxation time. Figure \ref{vis} shows the result of integrating
Eq. (\ref{eta}), where the $x$ axis represents different values of the
upper limit $t_{up}$. The total simulation time is divided in four
different runs and from them we obtain the error bars. Simulation
parameters are as indicated in Fig. \ref{pressureT}. The result for
this case is $\eta= 3.53(7) \times 10^{-3} $ fm $^{-3}$ when $t_{up}=
5000$ fm/c was chosen.

\subsection{Transport Cross Sections and Coefficients}
\label{Chapman-Enskog}

We obtain semi-analytical transport coefficients, for both quantum and
classical systems, by calculating the appropriate cross sections. Our
calculation follows the Chapman-Enskog formalism described in
Ref. \cite{gasestheory}. The quantity of interest is the transport
cross section,
\begin{equation}
  \phi^{(n)}=2 \pi
  \int_{-1}^{+1}d\cos\theta(1-\cos^n\theta)
\left.\frac{d\sigma(k,\theta)}{d\Omega}\right|_{c.m.},
\label{tr_crosssec}
\end{equation}
where the scattering angle $\theta$ and the collisional differential
cross-section $\frac{d\sigma(k,\theta)}{d\Omega}\bigr|_{c.m.}$ are
calculated in the center of mass reference frame of the two colliding
particles with momentum $\hbar k$. For indistinguishable particles, an
expansion of the cross-section in partial waves $\sum_{l} (2l+1)(e^{i
2 \delta_l}-1) P_l(\cos\theta)$ and the orthogonality of the Legendre
polynomials $P_l$ simplifies the integrals above to the infinite sums
\begin{eqnarray}
q^{(1)}\equiv\frac{\phi^{(1)}}{4\pi
a^2}&=&\frac{2}{x^2}{\sum_{l}}'(2l+1)\sin^2(\delta_l(x)),
 \label{tr_crosssec_sum1} \\
q^{(2)}\equiv\frac{\phi^{(2)}}{4\pi
a^2}&=&\frac{2}{x^2}{\sum_{l}}'\frac{(l+1)(l+2)}{(2l+3)}
\sin^2(\delta_{l+2}(x)-\delta_l(x)), \nonumber
 \label{tr_crosssec_sum2} \\ 
\end{eqnarray}
where the prime on the summation sign indicates the use of even $l$
for Bosons and odd $l$ for Fermions; $x=k a$ is a dimensionless
momentum variable with $a$ being the characteristic length scale of
the potential.  The quantity $4 \pi a^2$ with $a=\lambda_e$ for the
Yukawa potential represents a normalizing cross section that renders
the transport cross sections dimensionless.

Since the interparticle distance limits the number of partial waves
for scattering in a two-body system, we also introduce density
dependent quantum transport cross sections by limiting the number of
terms in the summation to
\begin{equation}
l_{n}=[k n^{-1/3}-1/2]\,, 
\label{lcutoff}
\end{equation}
where $n$ is the number
density and the quantity [c] denotes the integer part of c.

If the particles possess spin $s$, then the properly symmetrized forms
 are:
\begin{equation}
\begin{array}{ll}
q^{(n)}_{(s)}=\frac{s+1}{2s+1}q^{(n)}_{Bose}+
\frac{s}{2s+1}q^{(n)}_{Fermi},& \text{for integer } s,\\
\\
q^{(n)}_{(s)}=\frac{s+1}{2s+1}q^{(n)}_{Fermi}+
\frac{s}{2s+1}q^{(n)}_{Bose},& \text{for half-integer } s.
\end{array}
 \label{qn_spin}
\end{equation}

From Eqs.~(\ref{tr_crosssec_sum1}) and (\ref{tr_crosssec_sum2}), we
note that phase shifts are the central physical input for the
transport cross sections.  Details of the calculation of the phase
shifts, particularly for collisions of nuclei with large atomic number
$Z$ which involve a large number of partial waves and thus a large
number of scattering phase shifts, are provided in the appendix.

\begin{center}
\begin{figure}[tb]
	\includegraphics[width=3in]{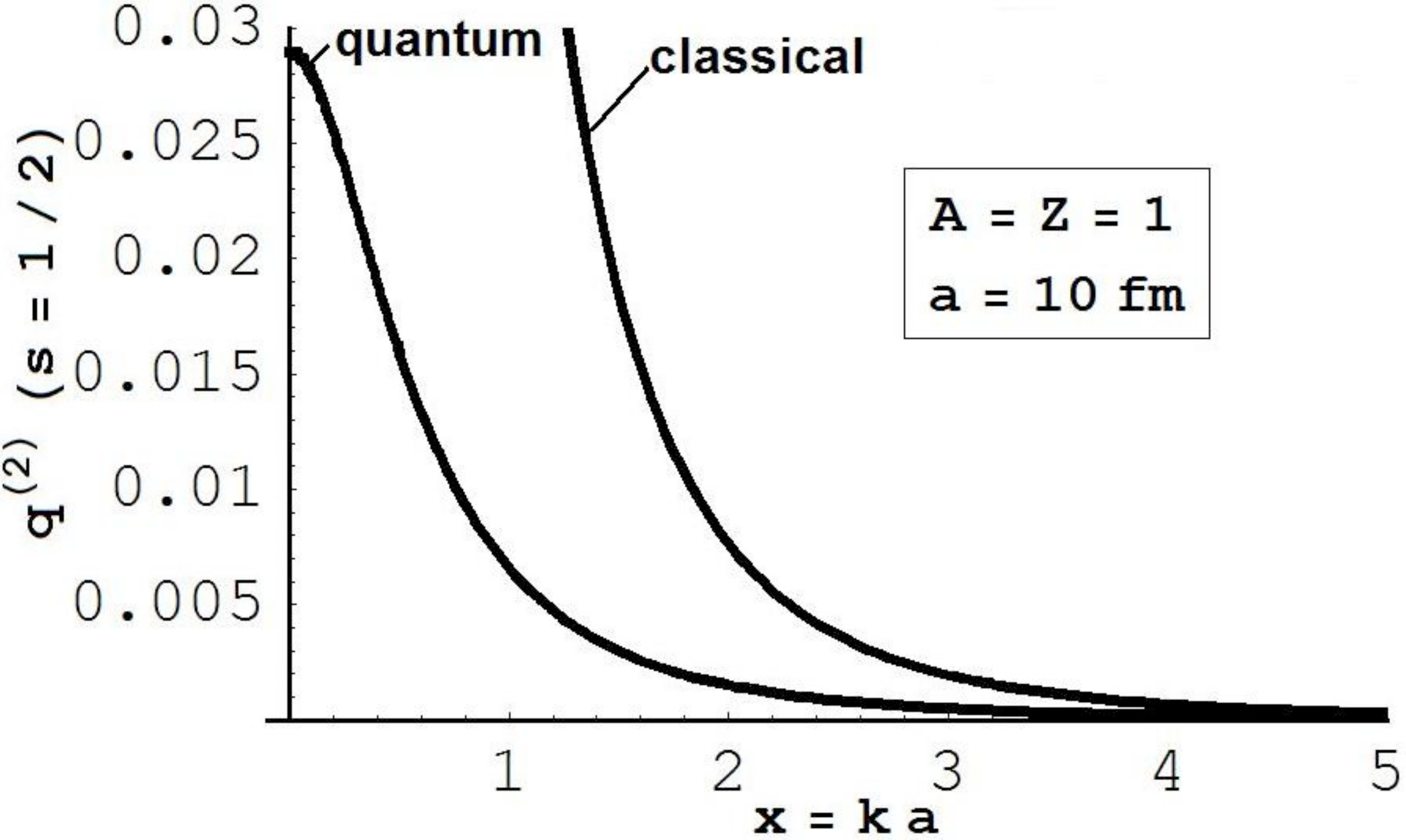}\\
	\includegraphics[width=3in]{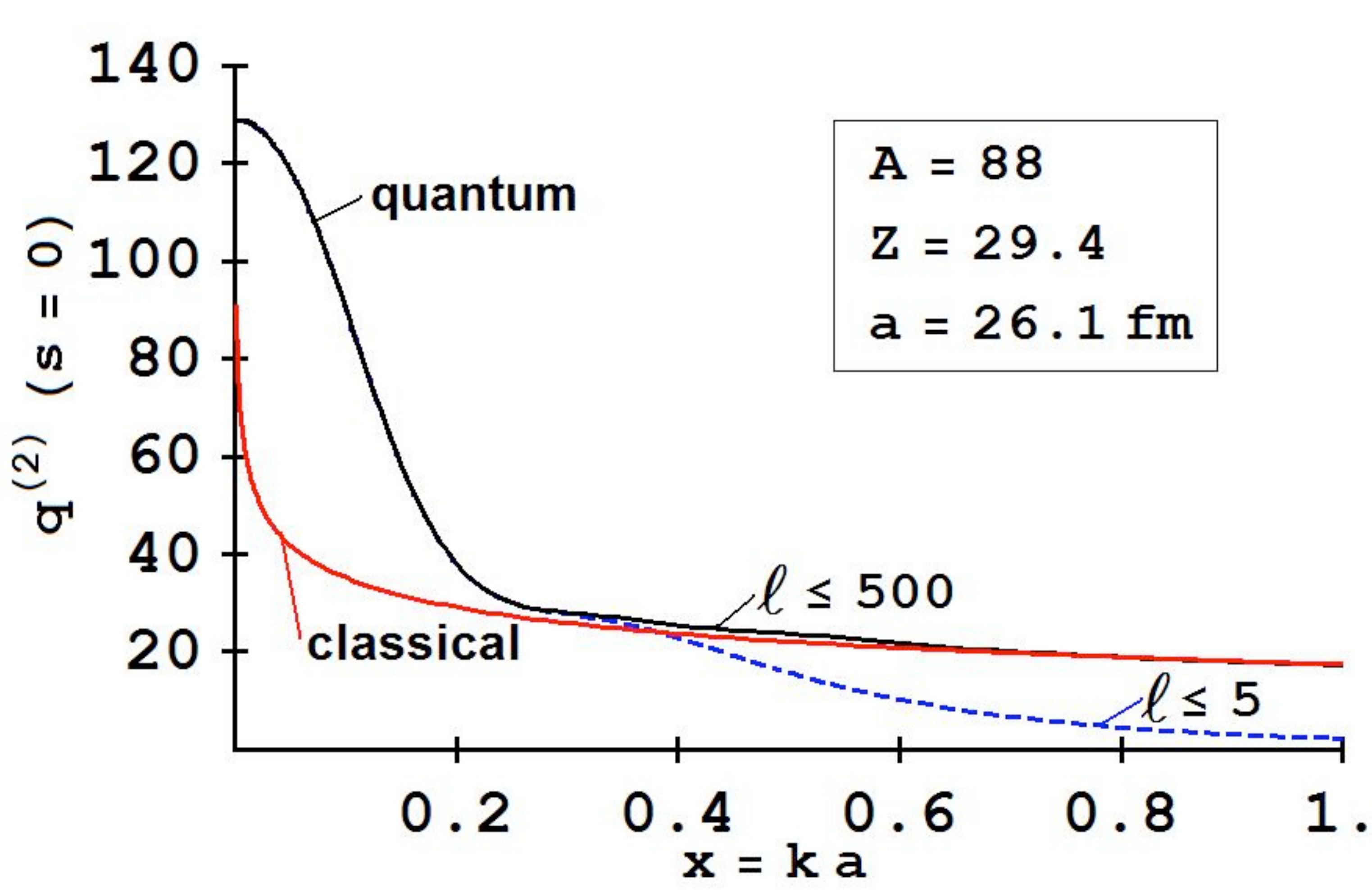}\\
\caption{(Color online.) Quantum and classical transport cross
  sections $q^{(2)}$ from Eqs.~(\ref{tr_crosssec_sum2}) and
  (\ref{eq_cl_qn}) at low energies for systems with $A=1$ and $A=88$
  in the dilute limit ($n\rightarrow 0$).}
	\label{Fig4_12}
\end{figure}
\end{center}

The classical transport cross section is given by
\begin{equation}
q^{(n)}=\frac{1}{4a^2}\int^\infty_0 \left(1-\cos^n(\theta) \right)db^2,
\label{eq_cl_qn}
\end{equation}
with $b$ the impact parameter, and 
\begin{equation}
  \theta(b,E) = \pi-2\phi_0 = 
\pi-2\int_{r_0(b,E)}^{\infty}
\frac{b \, dr}{r^2 \sqrt{\left(1-\frac{b^2}{r^2}-\frac{V(r)}{E}\right)}}
\label{eq_def_ang}
\end{equation}
the deflection angle which is a function of the energy $E$, $b$, and
$V(r)$. The quantity $r_0(b,E)$ is the distance of closest approach.
We also study the case in which the upper limit for $b$ in the
integral above is set equal to the average interparticle distance
$\approx n^{-\frac{1}{3}}$.  Note that this procedure introduces a
density dependence into the classical transport cross sections.

In the dilute limit, $n \rightarrow 0$, the classical and quantum
transport cross sections $q^{(2)}$ for systems with $A=88$ and $A=1$
are shown in Fig. \ref{Fig4_12}. As the energy tends to zero (or $x\to
0$), the quantum cross section stays finite whereas the classical one
diverges. For this reason, as we will see, the quantum result for
shear viscosity will differ from the classical result only for very
small temperatures.  For increasing $x$, the summation over $l$ in
Eq.~(\ref{tr_crosssec_sum2}) in the quantum case leads to results that
coincide with those of the classical case obtained using the
deflection function (see Eq.~(\ref{eq_cl_qn})).  For the system with
$A=88$, keeping only five partial waves ($l=0-5$) is adequate in the
energy region where $x\leq0.25$.

The transport coefficients can be calculated from the omega-integrals
\begin{equation}
\omega_\alpha^{(m,t)}(T)\equiv \int^\infty_0 d\gamma~ 
e^{-\alpha\gamma^2}\gamma^{2t+3}q^{(m)}(x),
\label{transport}
\end{equation}
where $\gamma=\frac{\hbar k}{\sqrt{2\mu k_BT}}$, $T$ is the
temperature, and $\mu$ is the reduced mass.  The formalism above
includes only binary collisions. Therefore, we expect it would be
accurate only at low densities.

In the first order (of deviations from the equilibrium distribution
function) approximation, the shear viscosity is given by~\cite{gasestheory}:
\begin{equation}
\frac{\left[\eta \right]_1}{\tilde{\eta}} = 
\left(\frac{a}{\lambda(T)}\right)\frac{1}{\omega_{1}^{(2,2)}(T)},
\label{visq}
\end{equation}
where $\lambda(T) = h/{\sqrt {2\pi Mk_BT}} $ is the
thermal de-Broglie wavelength. The quantity
\begin{equation}
\tilde{\eta} =\frac{5}{16\sqrt{2}}\frac{\hbar}{a^3}\,, 
\label{visqstar}
\end{equation}
is a characteristic viscosity of the Yukawa potential with
$a=\lambda_e$.  For later use, we define here the
characteristic temperature $k_B\tilde{T}\equiv\frac{2\pi \hbar^2}{a^2
M}$ (or $\frac{T}{\tilde{T}}=\left(\frac{a}{\lambda}\right)^2$).

Equation (\ref{visq}) shows clearly that if $\omega_1^{(2,2)}$ is
$T$-independent (as for rigid-spheres with a constant cross section),
the shear viscosity exhibits a $T^{1/2}$ dependence which arises
solely from its inverse dependence with $\lambda (T)$. For
energy-dependent cross sections, however, the temperature dependence
of the viscosity is sensitive also to the temperature dependence of
the omega-integral.

In the second order approximation~\cite{Uehling}
\begin{equation}
\frac{\left[ \eta\right]_2}{\left[ \eta\right]_1} =
\left(1+\delta_{\eta}(T)\right)\left(1\pm n \lambda^3
\epsilon_{\eta}(T)\right),
\label{Diff2}
\end{equation}
where $\pm$ means plus sign for Bose and minus for Fermi statistics,
$n$ is the number density,
\begin{equation}
\delta_{\eta}\equiv \frac{3(7\omega_{1}^{(2,2)}-2\omega_{1}^{(2,3)})^2}
{2\left(\omega_{1}^{(2,2)}\left(77\omega_{1}^{(2,2)} + 
6\omega_{1}^{(2,4)}\right)-6\left(\omega_{1}^{(2,3)}\right)^2\right)},
\label{dvisq2}
\end{equation}
and
\begin{equation}
\epsilon_{\eta}\equiv 2^{-7/2} 
\left[4-\frac{128}{3^{3/2}}\frac{\omega_{4/3}^{(2,2)}}
{\omega_{1}^{(2,2)}}\right].
\label{epseta}
\end{equation}

It is worthwhile to note that at the first order of deviations from
the equilibrium distribution function, the viscosity is independent of
density, unless density dependent cut-offs are used to delimit the
quantum or classical transport cross sections. An explicit density
dependence arises only at the second order.

\section{MD Simulations of a Dilute Plasma}
\label{dilute}
We calculate the shear viscosity for a dilute plasma following the
procedure described in Secs. \ref{sec:IIIM} and
\ref{sec:autoc}. The results presented here correspond to Hydrogen,
$Z=1$ and $A=1$. The box length of the simulation volume for the
parameters used is $L \sim$ 100 fm. To minimize finite size effects in
our MD simulations, it is needed that $L\gg \lambda_e$, where
$\lambda_e$ is the screening length of the Yukawa interaction, see
Eq. (\ref{potential}).  To assure this condition, we arbitrarily fix the
electron screening length $\lambda_e$ to $10$ fm. Also, we choose
values of density and temperature such that the system is weakly
coupled, $\Gamma<1$.

In table \ref{tabledilute} we summarize the values of the ion density
$n$, Coulomb parameter $\Gamma$, and simulation times used. The time
step $\Delta t_{MD}$ is 10 fm/c.  
We will use the results from these MD simulations here and later in 
Sec. \ref{quantum} to compare with results obtained from calculations
performed using the material in Sec. \ref{Chapman-Enskog}.

Here, we point out that our numerical simulations in the limit when $n
\rightarrow 0$ are involved. To follow the trajectories of a highly
dilute gas implies very small simulation time steps, and long
simulation times in order to find correlations between the
ions.

\begin{table}
\begin{center}
\caption{Simulation runs at $T$ = 0.1 MeV for a H dilute
plasma. $N_{ion}$ is the total number of ions in the system. The time
step in the MD simulation is $\Delta t_{MD}=10$ fm/c. $T_W$ is the
warm up time, $T_M$ the measurement time, and $\Delta t_{ P_{xyz}}$ is
the time step used to calculate the pressure tensor.  }
\begin{tabular}{llllllll}
\hline
 $N_{ion}$ &$\Gamma$& $n (\times 10^{-3}$fm$^{-3}$)  & $T_W
 (10^5$fm/c)  
& $T_M (10^6 $fm/c)  & $\Delta t_{P_{xyz}}$ \\
%     &           &        &  fm/c & fm$^{-3})$ & fm/c) &  fm/c) & (MeV) \\   s
 500&0.5&0.01&-&8.0&10\\
 100& 1.07&$0.1$ & $21.9$ & $97.0$ & 100\\
 100& 1.46&$0.25$ & $0.2$ & $100$ & 100 \\
 100  & 1.84& $0.5$ & $0.2$ & $97.0$ & 100\\
 500 & 2.32& $1.0$ & $47.2$ & $12.5$ & 10 \\ 
 500 & 3.15& $2.5$ & $1.3$ & $10$ & 10 \\

\hline
\end{tabular} 
\label{tabledilute}
\end{center}
\end{table}

Table \ref{tabledilute2} shows results for the viscosity for different
values of temperature and density. Again we use $A=1$, and $Z=1$. The
three first rows correspond to cases in which the density was kept
constant. The viscosity is a monotonically increasing function of $T$.
We find from our simulations that the dilute plasma behaves as a gas
where higher temperatures provide higher momenta, and hence lead to
larger momentum flux.

Table \ref{tabledilute2} also shows the feature of increasing viscosities
for large densities.  As the temperature decreases the differences due
to changes in density are more evident. We expect that as the
temperature decreases the changes in the dependency of $\eta$ vs $n$
to be more dramatic making the plot steeper due to the effect of
correlations between ions. For example, $\eta$ remains in the same
order of magnitude $\backsim 10^{-4}$, when $n$ changes by two orders
of magnitude. On the other hand,
comparing the runs at $T=0.05$ MeV and $T=0.01$ MeV (close values) in
Table \ref{tabledilute2}, we observe a change in $\eta$ of two orders
of magnitude, when $n$ changes in the same way. This is due to the
fact that there is a larger change in the inter-ion distance, which
increases the correlations between ions, than the change in the ratio
of thermal to Coulomb energy $\Gamma$.

\begin{table}
\begin{center}
\caption{(Color online.) Viscosity results (given by setting $\hbar
  =1$) for a H dilute plasma and
parameters used in the simulations. $N_{ion}$, the total number of
ions in the system, is 500. The time step in the MD simulation is
$\Delta t_{MD}$, whereas $T_{MD}$ is the total simulation time, and
$\Delta t_{ P_{xyz}}$ is the time step used to calculate the pressure
tensor.  }
\begin{tabular}{lllllll}
\hline
 $T$& $n$ &$\Gamma$  &$\eta $  & $T_{MD}  $& $\Delta
 t_{P_{xyz}}$ 
& $\Delta t_{MD}  $  \\
 (MeV)&(fm$^{-3}$)&&(fm$^{-3}$)&($10^7 $fm/c) &&(fm/c) \\
%     &           &        &  fm/c & fm$^{-3})$ & fm/c) &  fm/c) & (MeV) \\   
 0.5   & 10$^{-3}$ & 0.46&$3.9(3) \times 10^{-3}$ & $7.35$ & 100&5  \\
  0.1   & 10$^{-3}$ &2.32 & $3.28(7) \times 10^{-4}$ & $1.72$ & 10 &10  \\ 
 0.05&10$^{-3} $&4.64 & $1.51(3) \times 10^{-4} $&1.08&10&10\\
0.01&10$^{-5}$&5.0&$7.0(2)\times 10^{-6}$&0.8&10 &10\\
0.001& 10$^{-5}$&50& $9.1(5)\times 10^{-7}$ &0.8 &10&10\\

\hline
\end{tabular} 
\label{tabledilute2}
\end{center}
\end{table}

\section{Dilute Limit Quantum and Classical Viscosities}
\label{quantum}

\begin{center}
\begin{figure}[tb]
	\includegraphics[width=3.5in,angle=0]{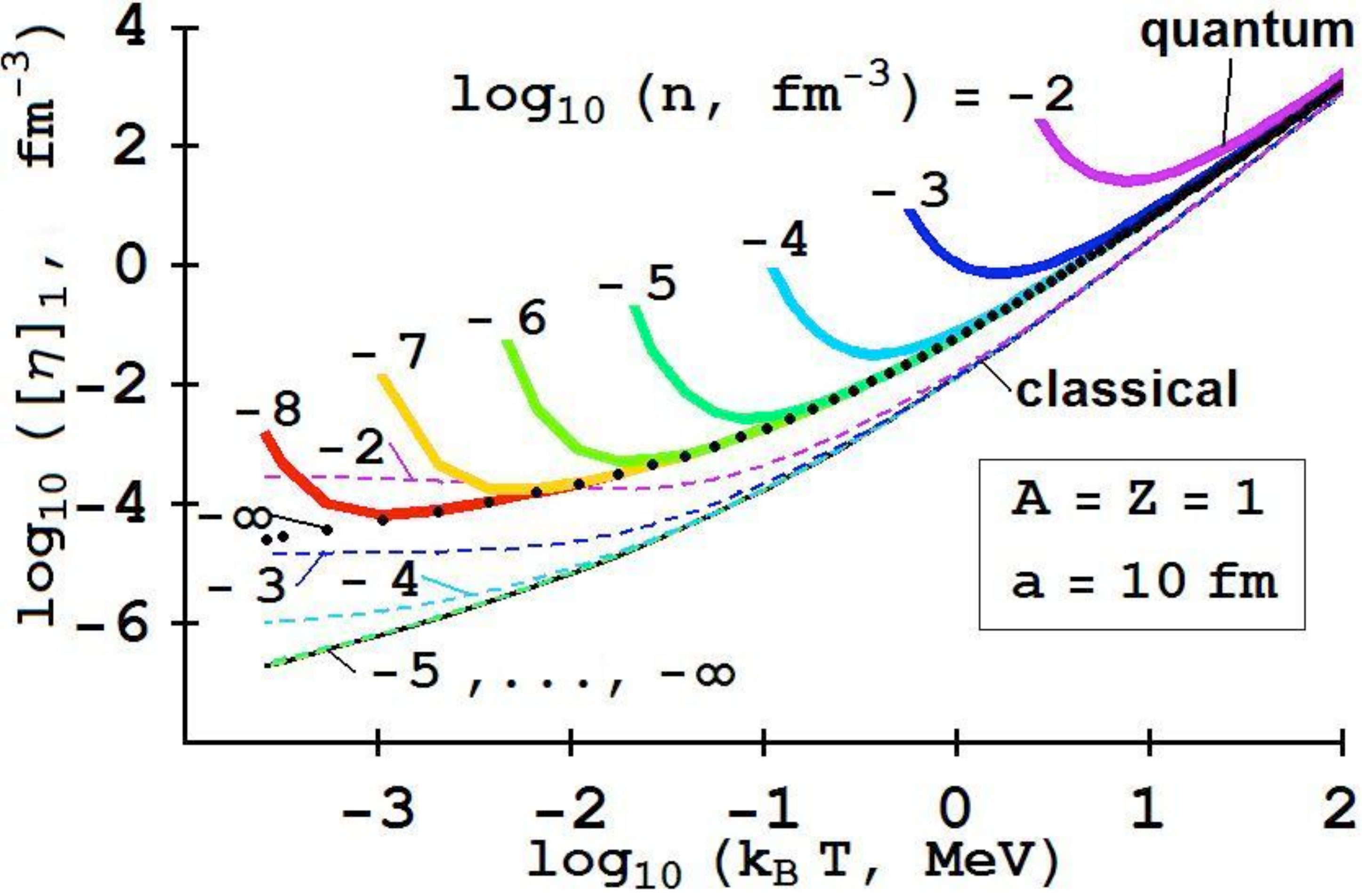}\\
\caption{(Color online.) Quantum and classical results from
Eq. (\ref{visq}) for the first order shear viscosity $[\eta]_1$ for a
H dilute plasma.  (Viscosities in this paper are given by setting
$\hbar = 1$). The lower-most solid curve is the classical dilute gas
limit. The dots show the corresponding limit in the quantum case. The
upward-bending solid (dashed) curves are results obtained using
density dependent cut-offs in the quantum (classical) case and signal
the onset of more than two-body effects.}
	\label{lgEta1lgkTclqnZ1}
\end{figure}
\end{center}

In this section we compare results based on the Chapman-Enskog
formalism outlined in Sec. \ref{Chapman-Enskog} with those from the MD
calculations in Sec. \ref{dilute}.  In Fig. \ref{lgEta1lgkTclqnZ1} the
quantum and classsical results from Eq. (\ref{visq}) for the first
order shear viscosity $[\eta]_1$ for the system with $A=1$ are shown
as functions of temperature.  As noted earlier, both classical (the
lower-most solid curve) and quantum (the dotted curve) viscosities are
independent of density at first order.  As expected, the quantum
results approach the classical ones at high temperature. With
decreasing temperature, however, the viscosity for the quantum case is
significantly larger than its classical counterpart due to the smaller
transport cross sections (see Fig.~\ref{Fig4_12}).

The upward-bending solid (quantum case) and dashed (classical case)
curves in this figure show results obtained by imposing the density
dependent cut-offs on the angular momentum in the quantum case,
Eq.~(\ref{lcutoff}), and the impact parameter in the classical
case. These results allow us to establish the ranges of density and
temperature for which the first order result is valid. For example, at
a density of $10^{-5}~{\rm fm}^{-3}$, the quantum results for
temperatures below $T \sim 0.1$ MeV are susceptible to more than
two-body effects (not considered in this treatment) so that the first
order result is not reliable.   
For the range of temperatures shown in
this figure, effects of the density dependent cut-offs are more
significant for the quantum case than for the classical case.
\begin{center}
\begin{figure}[htbp]
\includegraphics[width=3.5in]  {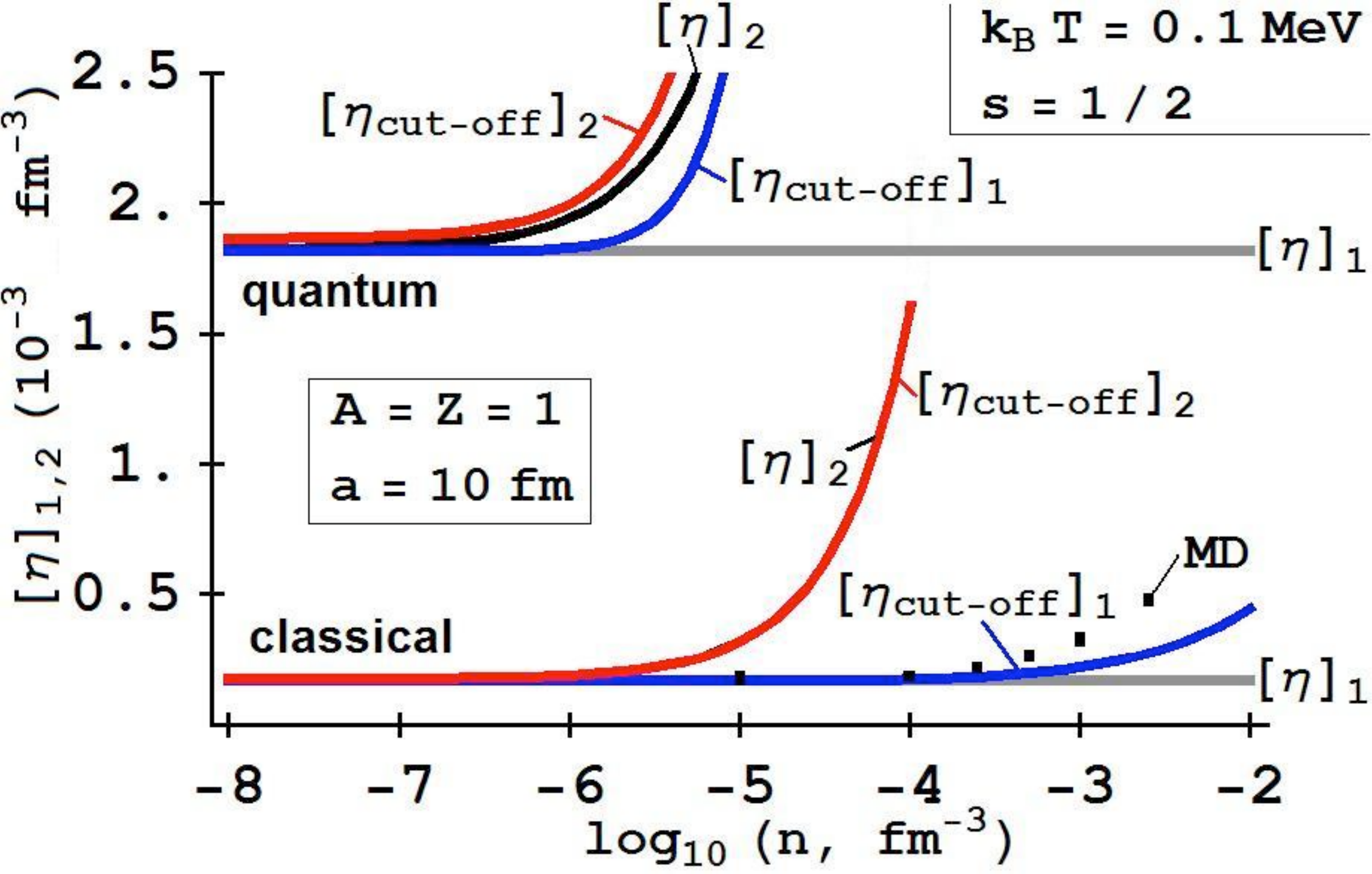}
\caption{(Color online.) Quantum and classical viscosities as a
function of the ion density $n$, for a H plasma. The temperature is
0.1 MeV.  First and second order results from Eqs.~(\ref{visq}) and
(\ref{Diff2}) are denoted by $[\eta]_1$ and $[\eta]_2$,
respectively. The filled squares are results of MD simulations.}
\label{vis_dense}
\end{figure}
\end{center}
For the system with $A=1$, results of $[\eta]$ up to second order from
Eqs.~(\ref{visq}) and (\ref{Diff2}) are shown in Fig.~\ref{vis_dense}
as functions of density. The temperature in this case is $0.1$ MeV.
The upward-bending curves show effects of density dependent cut-offs.
Results from the MD calculations are shown as filled
squares. Noteworthy features in this figure are: (1) the large
differences between the quantum and classical results, and (2) the
disposition of the MD results with respect to the classical results.
These results also underscore the importance of incorporating quantum
effects in MD calculations, even at low densities.

In order to understand these results, we begin by noting that the
characteristic shear viscosity and temperature in this case are
$\tilde{\eta}=2.2\times 10^{-4}~{\rm fm}^{-3}$ (we have set $\hbar =1$) and
$k_B\tilde{T}=2.6$ MeV. For $k_BT = 0.1$ MeV and $A=1$, we find
$[\eta]_1~(classical) = 1.72\times 10^{-4}~{\rm fm}^{-3}$
and $[\eta]_1~(quantum) = 1.82\times 10^{-3}~{\rm fm}^{-3}$.

The fact that the first order viscosity in the quantum case is nearly
ten times larger than the classical result can be understood by
examining the ratio of the corresponding omega-integrals in
Eq.~(\ref{transport}). In the quantum case, the integrand of the
omega-integral peaks at $\gamma \simeq 2$ for which we find $x_{peak}
\simeq 2 {\sqrt {2\pi T/\tilde T}} \sim 1$.  At this value of
$x_{peak}$, the quantum transport cross section is nearly ten times
smaller than the classical one (see Fig.~\ref{Fig4_12}) which
quantitatively accounts for the ratio of $\sim 10$ being sought.  In
physical terms, the differences in viscosities stem from the
differences in the transport cross sections: in the quantum case, the
cross section saturates at low energies whereas in the classical case
the cross section diverges. 

The densities at which the viscosities begin to be dependent on
angular momentum or impact parameter cut-offs are also evident from
Fig.~\ref{vis_dense}.  In the classical case, effects of the density
dependent cut-off enter at much higher densities than do differences
stemming from going to a higher order (compare $[\eta]_1$ and
$[\eta]_2$).  It is worthwhile to mention that when these corrections
are significant, many-body correlations not considered here will be
important. This physical effect is amply demonstrated by the results
of the classical MD calculations which lie between the classical
values of $[\eta]_1$ and $[\eta]_2$.

\begin{center}
\begin{figure}[htbp]
\includegraphics[width=3.5in]  {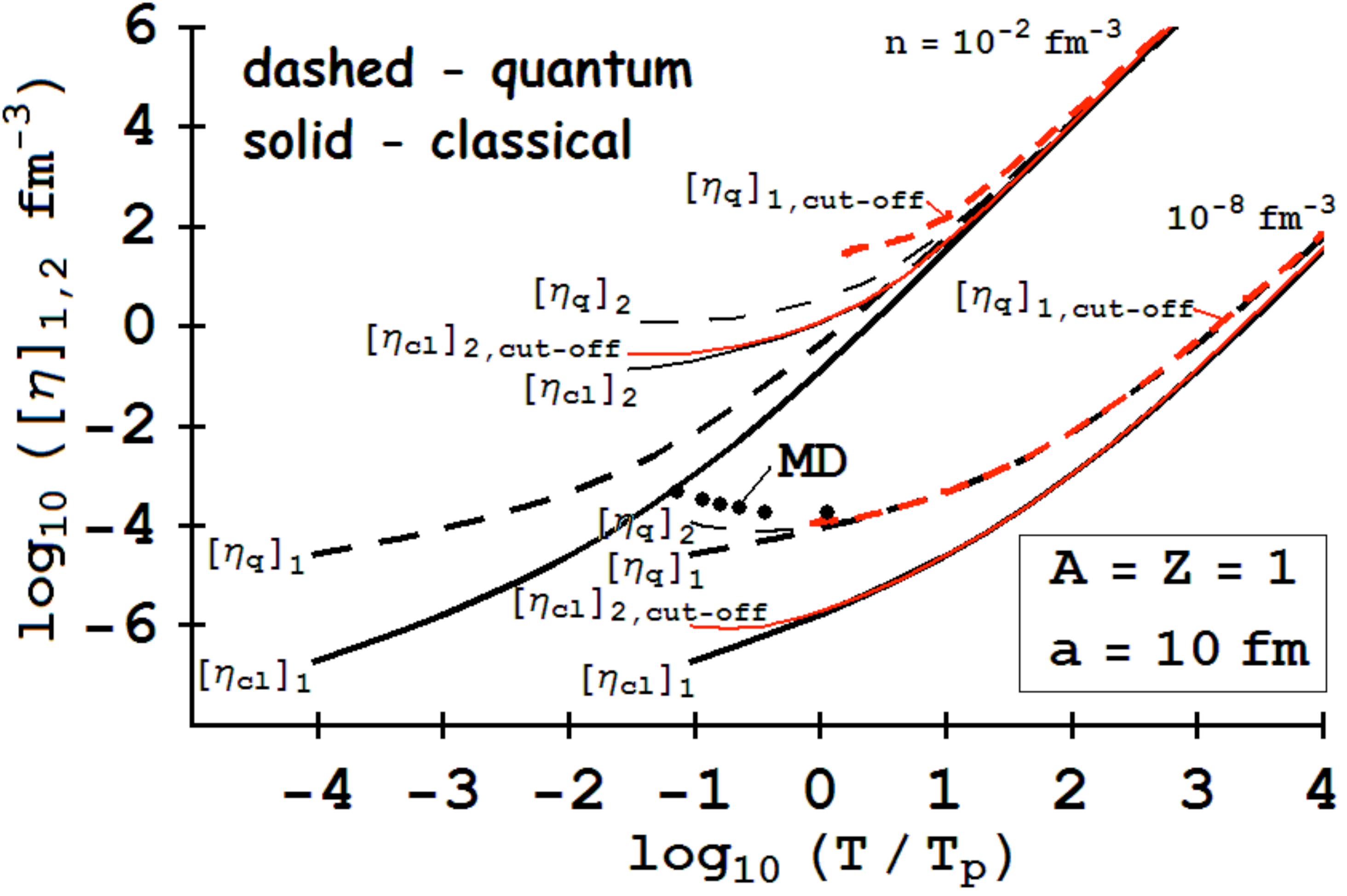}
\caption{(Color online.) Quantum and classical viscosities as a
function of $T/T_p$ for a H plasma. The plasma temperature $T_p$ is
given in Eq.~(\ref{plasmaT}).  The densities are as indicated.  First
and second order results from Eqs.~(\ref{visq}) and (\ref{Diff2}) are
denoted by $[\eta]_1$ and $[\eta]_2$, respectively.}
\label{TOTP1}
\end{figure}
\end{center}

In order to appreciate how and when quantum effects become important in a
Hydrogen plasma, we show in Fig.~\ref{TOTP1} results of the dilute
limit quantum and classical first and second order viscosities for two
densities as functions of the ratio $T/T_p$, where $T_p=\omega_p
\simeq 27.5~(Z^2n/A)^{1/2}$ MeV (with $n$ in fm$^{-3}$) is the plasma
temperature in Eq.~(\ref{plasmaT}).  Results obtained with
density-dependent cut-offs are also included in this figure to
illustrate when three-particle effects become important.  Results for
the intermediate densities (not shown in the figure for the sake of
clarity) are quantitatively different, but qualitatively similar.  The
filled circles are the results of MD simulations 
corresponding to a temperature of 0.1 MeV at the densities shown in
Fig.~\ref{vis_dense}. The trajectory traced by the MD results is due
to the density dependence of $T_p$.

The upward shift in the magnitudes of the viscosities (which are
intrinsically independent of density) with increasing density in
Fig.~\ref{TOTP1} is caused by the $n^{1/2}$ dependence of the plasma
temperature $T_p$.  Quantum effects lead to viscosities that are
significantly larger than the classical results as $T/T_p$ decreases;
the lower the densities the larger are the differences. Put
differently, the values of $T/T_p$ for which the quantum and classical
results merge together increase as the density decreases. It must be
borne in mind, however, that although the quantum effects are captured
fully in the cross sections the thermal weightings are classical in
our treatment here. With increasing density, effects of the Pauli
principle and possible in-medium and many-body effects will become
important. As shown by the second-order viscosities and the results
obtained using density-dependent cut-offs, more than two-body effects
become important when $\log_{10}(T/T_p) \simeq 0~(2)$ for $n=10^{-8}~
(10^{-2})~{\rm fm}^{-3}$.

\begin{center}
\begin{figure}[tb]
	\includegraphics[width=3.5in,angle=0]{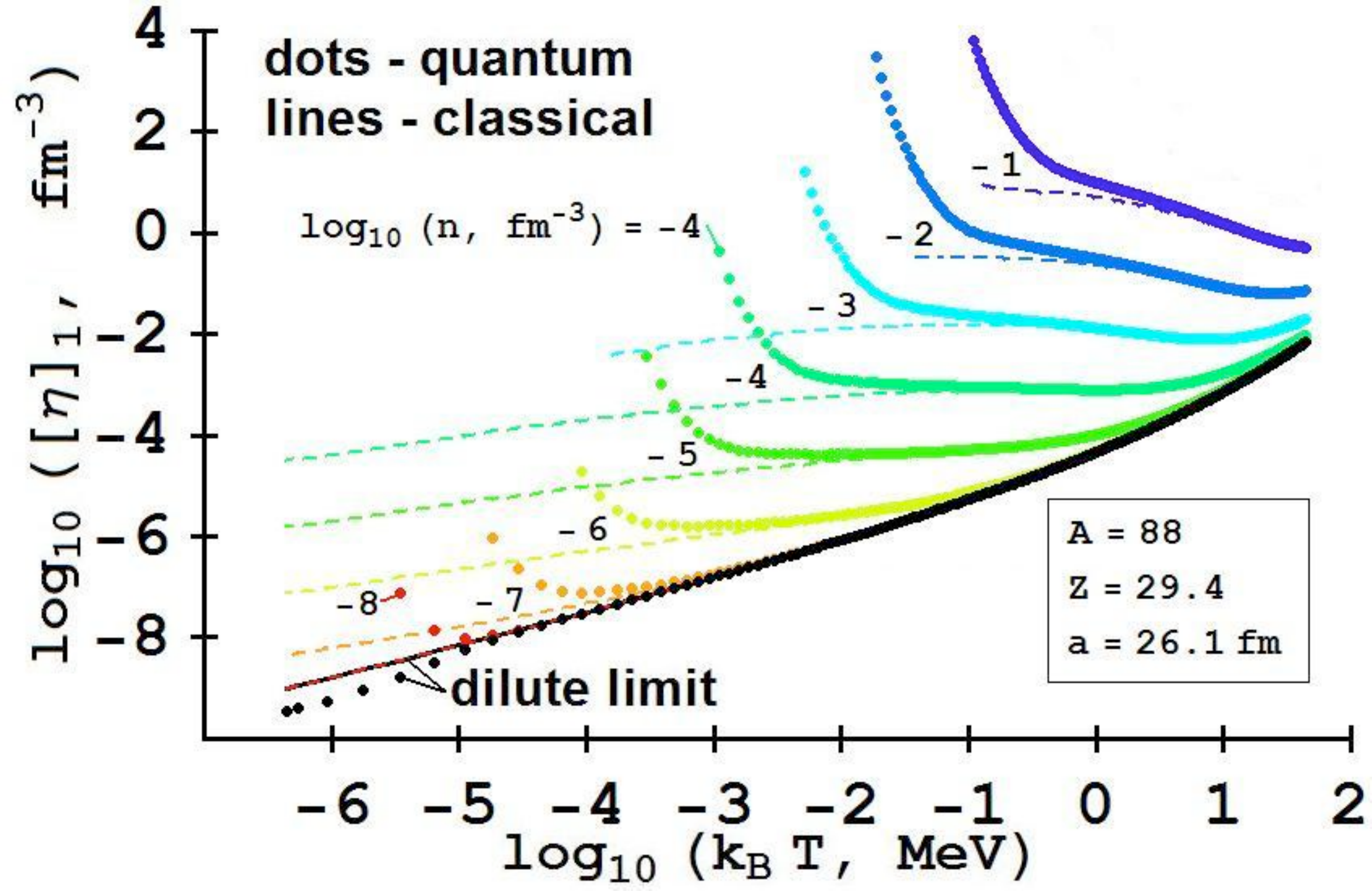}\\
\caption{(Color online.) 
Same as Fig.~\ref{lgEta1lgkTclqnZ1}, but for the system with $A=88$.}
	\label{figEta1kTqcl}
\end{figure}
\end{center}

The result for shear viscosity at first order is shown in
Fig. \ref{figEta1kTqcl} for the system with A=88.  As the density
decreases, the classical (lines) and quantum (dots) results merge
together as expected in the dilute limit.  Differences between the two
cases occur only as the temperature decreases to values below the
temperature corresponding to the plasma temperature.  The
characteristic shear viscosity and temperature in this case are
$\tilde{\eta}=1.2\times 10^{-5}~{\rm fm}^{-3}$ and $k_B\tilde{T}=4.4
\times 10^{-3}$ MeV.  For $k_BT = 1$ MeV and $A=88$, we find
$[\eta]_1~(classical) = 4.63\times 10^{-5}~{\rm fm}^{-3}$ and
$[\eta]_1~(quantum) = 4.62\times 10^{-5}~{\rm fm}^{-3}$.

\begin{center}
\begin{figure}[tb]
	\includegraphics[width=3.5in,angle=0]{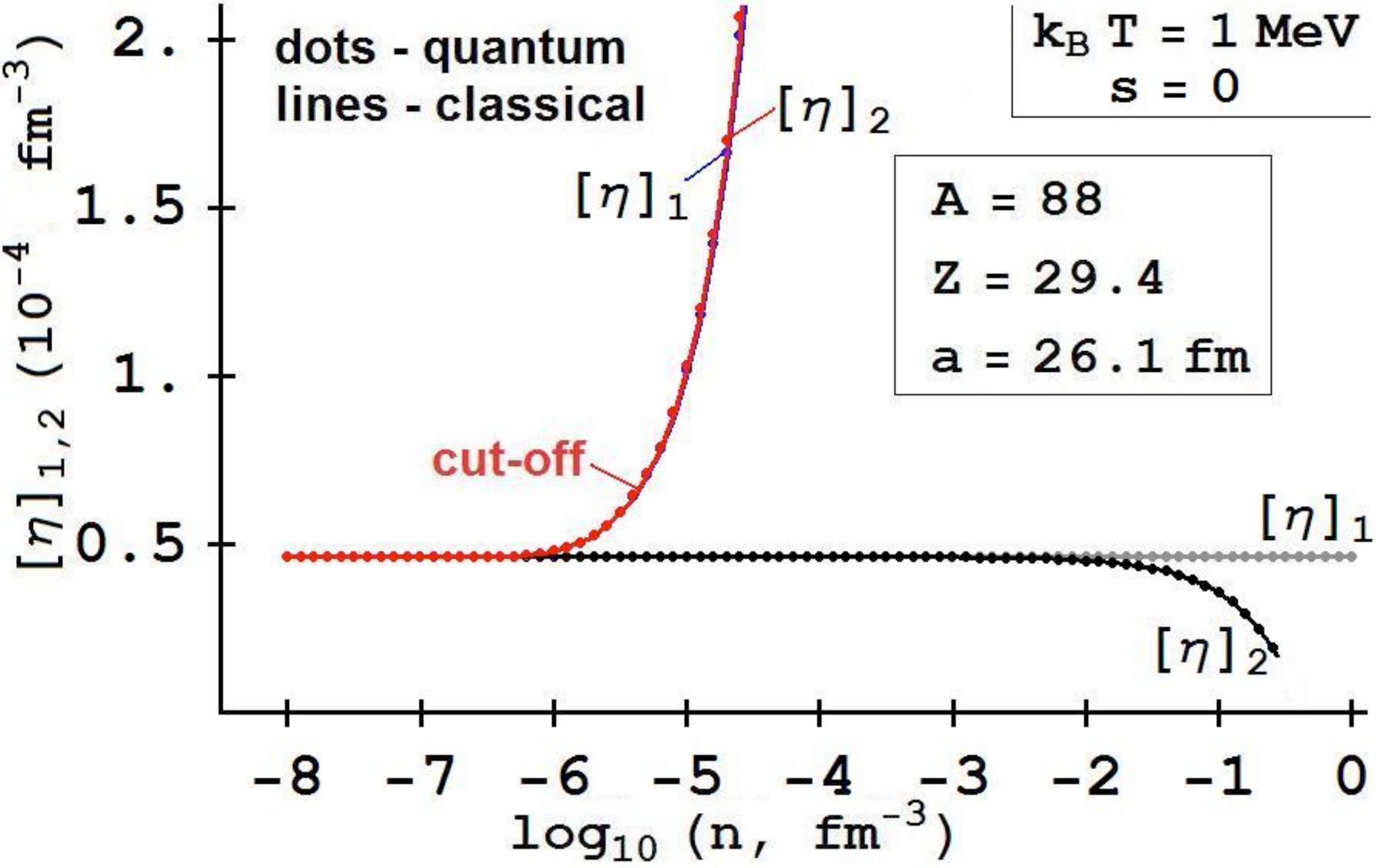}\\
\caption{(Color online.) 
Same as Fig.~\ref{vis_dense}, but for the system with $A=88$.} 
	\label{Eta12lgnZ29}
\end{figure}
\end{center}

Results of  $[\eta]$ up to second order are shown in
Fig. \ref{Eta12lgnZ29} for the system with 
$A=88$ as functions of density at a temperature
of $1$ MeV.  There is virtually no
difference between the classical and quantum results because the
system is essentially classical. Effects of the density dependent cut-off
enter at much lower densities than do differences stemming from going to a
higher order (compare $[\eta]_1$ and $[\eta]_2$).

\begin{center}
\begin{figure}[htbp]
\includegraphics[width=3.5in]  {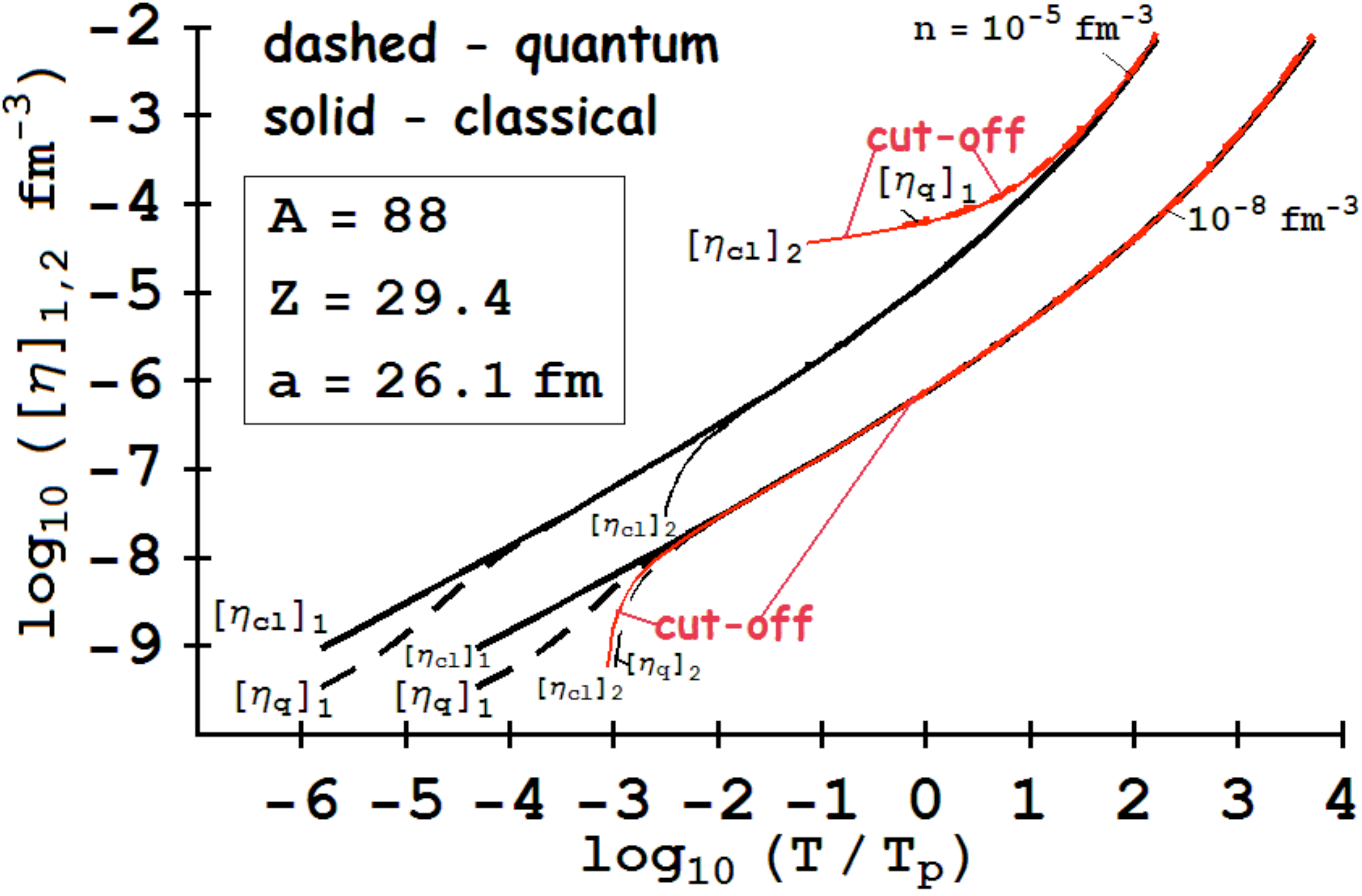}
\caption{(Color online.) Same as Fig.~\ref{TOTP1}, but for the
  heavy-ion plasma with A=88. }
\label{TOTP88}
\end{figure}
\end{center}

Figure \ref{TOTP88} shows how the viscosities of the heavy-ion plasma
with $A=88$ depend on the ratio $T/T_p$. Except for $T/T_p \ll 1$,
the classical and quantum results are indistinguishable. The values of
$T/T_p$ for which more than two-body effects are important are
easily discerned from this graph by inspecting the results obtained
using density-dependent cut-offs.

\subsection{Comparison of Dilute Limit Viscosities with MD Results}

We find that MD results for the Hydrogen plasma are
in good agreement with the classical semi-analytical calculations at
low densities. On the other hand, the first-order semi-analytical
solution predicts a constant behavior of the viscosity for all values
of the ion density.  MD simulations have taken into account
correlations between ions which have not been included in the
semi-analytical results. These correlations have a stronger effect as
density increases leading to a larger difference from the
semi-analytical description. From our results we conclude that the
viscosity for a dilute H plasma is constant at low densities and
rapidly increases for higher densities. Results for $T\ll T_P$, with
$T_P$ the plasma frequency, could involve quantum corrections and we
expect our MD results to be inaccurate in that regime.

We emphasize the agreement between our MD simulations and
semi-analytic results in the low density classical limit.  This
provides a significant check for both approaches.  As the density
increases, MD simulations fully include correlations between ions and
are therefore directly applicable at any density.  In contrast the
semi-analytic approach only works at low densities.  The Yukawa
interaction is reasonably long ranged.  It can still be important at
distances of a few or more screening lengths $\lambda_e$.
Furthermore, there are a large number of other ions to interact with
at large distances.  This limits the applicability of the
semi-analytic approach to low densities.

On the other hand, the quantum semi-analytical viscosities for the
Hydrogen plasma differ from those of classical semi-anlytical and MD
simulations by about an order of magnitude. This large difference
points to the need for including quantum effects in MD simulations of
light ions. However, see results in Sec.~\ref{visle} in which density
dependent screening lengths are used.

The fact that the MD result $\eta= 3.53 \times 10^{-3} $ fm $^{-3}$
for the system with $A=88$ at $k_BT=1$ MeV and $n=7.18\times
10^{-5}~{\rm fm}^{-3}$ is about two orders of magnitude larger than
the semi-analyical result of $4.63\times 10^{-5}~{\rm fm}^{-3}$
deserves some comments. From Fig. \ref{figEta1kTqcl}, we note
that at this density the cut-off dependence (which signals the
influence of more than two-body effects) in the semi-analtical
viscosities sets in at a temperature of about 10 MeV.  Alternatively,
Fig. \ref{Eta12lgnZ29} shows that at a temperature of 1 MeV,
the dilute limit is reached only below a density of $10^{-6}~{\rm
fm}^{-3}$.  Thus for the MD results to approach the semi-analytical
results, either an order of magnitude higher temperature is required
at this density or an order of magnitude lower density is needed at
this temperature. For the values of density and temperature chosen,
many-body effects will be important.  Note that the second-order
semi-analytical approach is an attempt, but only in a perturbative
fashion (as an expansion in the parameter $(na^3)$). For this
perturbative result to be trustable, it should not exceed the
first-order result substantially.  The MD calculations take into
account many-body effects to all orders, albeit classically in our
calculations. For heavy-ion systems in the dilute limit, quantum
effects are not expected to play an important role because a lot of
partial waves contribute.

\subsection{Dilute Limit Viscosities for Density Dependent Screening Lengths}
\label{visle}
\begin{center}
\begin{figure}[tb]
	\includegraphics[width=3in]{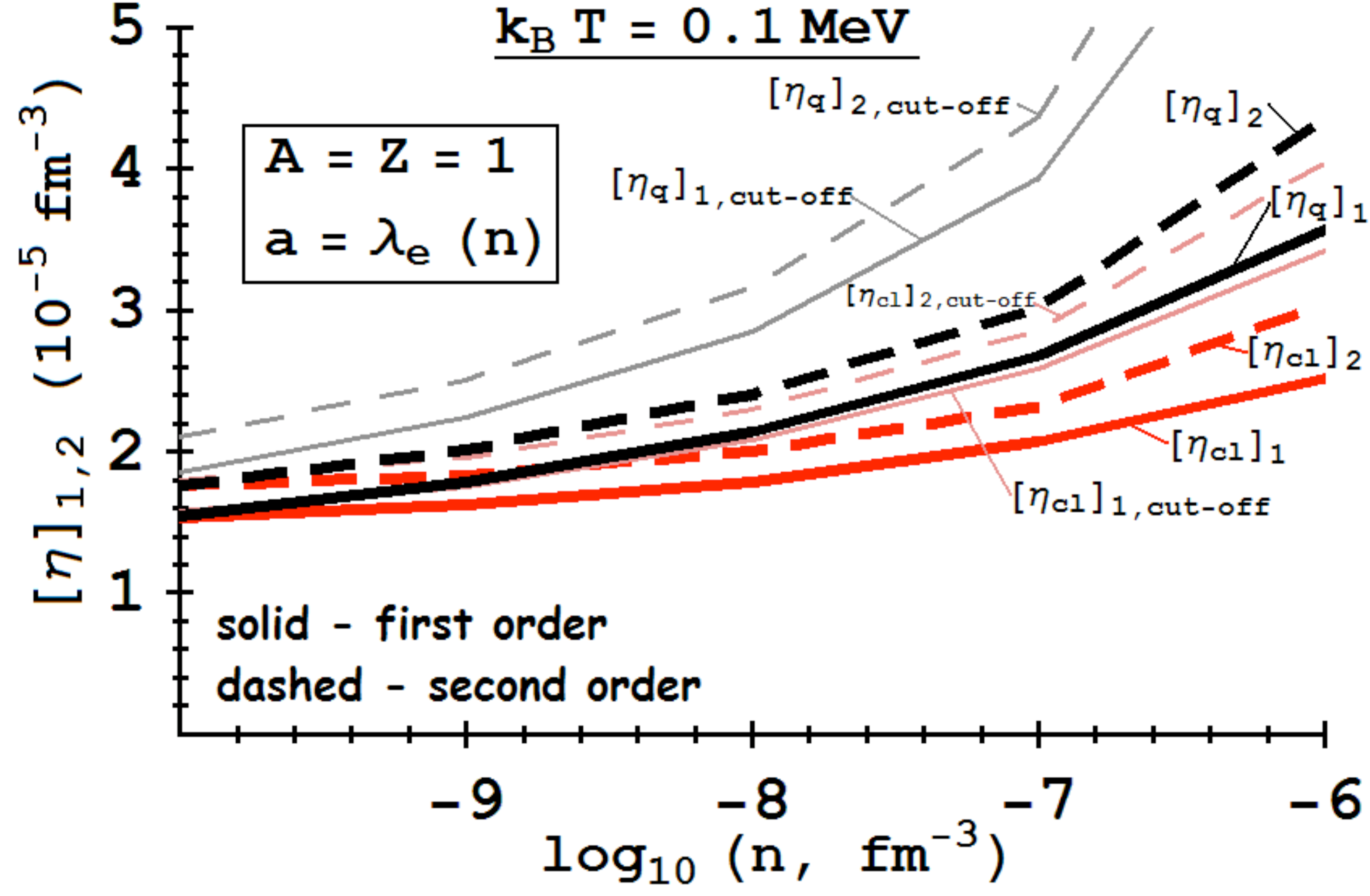}\\
\caption{(Color online.) Viscosities for density dependent screening
  lengths as functions of density.}
	\label{one}
\end{figure}
\end{center}

The electron screening in length in a charge neutral plasma is given by 
\begin{eqnarray}
\lambda_e =\frac {\pi^{1/2}}{2ek_F} \cong
3.35~\left(\frac {1}{\langle Z\rangle n} \right)^{1/3} \,. 
\label{lambdae}
\end{eqnarray}
For example, at a density of $10^{-6}~{\rm fm}^{-3}$, $\lambda_e
\simeq 336~(108)$ fm for $\langle Z \rangle = 1~(29.4)$. As MD
calculations are time-wise prohibitive for very large screening
lengths, we report here on the dilute limit semi-analytic calculations
of Sec.~\ref{Chapman-Enskog} in which the parameter $a$ in the Yukawa potential
describing the ion-ion interaction is set equal to the density
dependent screening length $\lambda_e(n)$.  As the screening length is
density dependent, the shear viscosity exhibits distinct features as a
function of density.  In Fig.~\ref{one}, results of viscosities for
the Hydrogen plasma are shown as a function of density.  Even at first
order the shear viscosity is density dependent through the potential
and increases with increasing $n$. Corrections arising from second
order deviations from the equilibrium distribution function contribute
significantly even at low densities.  Results obtained by imposing
density dependent cut-offs indicate that more than two-body physics
plays an important role at low densities in both the classical and
quantum cases.  The large screening lengths at low densities induce a
large number (up to 1000) of partial waves to contribute in the
quantum case. For moderate densities, however, significantly lower
number of partial waves are needed to obtain convergent results.  Both
the classical and quantum results are nearly the same at low
densities. However, the quantum results are larger than the classical
ones as the density increases because the classical cross section is
larger for a smaller screening length at a fixed energy ($x=ka$ gets
smaller). Note that imposing density cut-offs on the classical results
mimics the quantum results without cut-offs.
 
%\vspace*{3cm}

\begin{center}
\begin{figure}[tb]
	\includegraphics[width=3in]{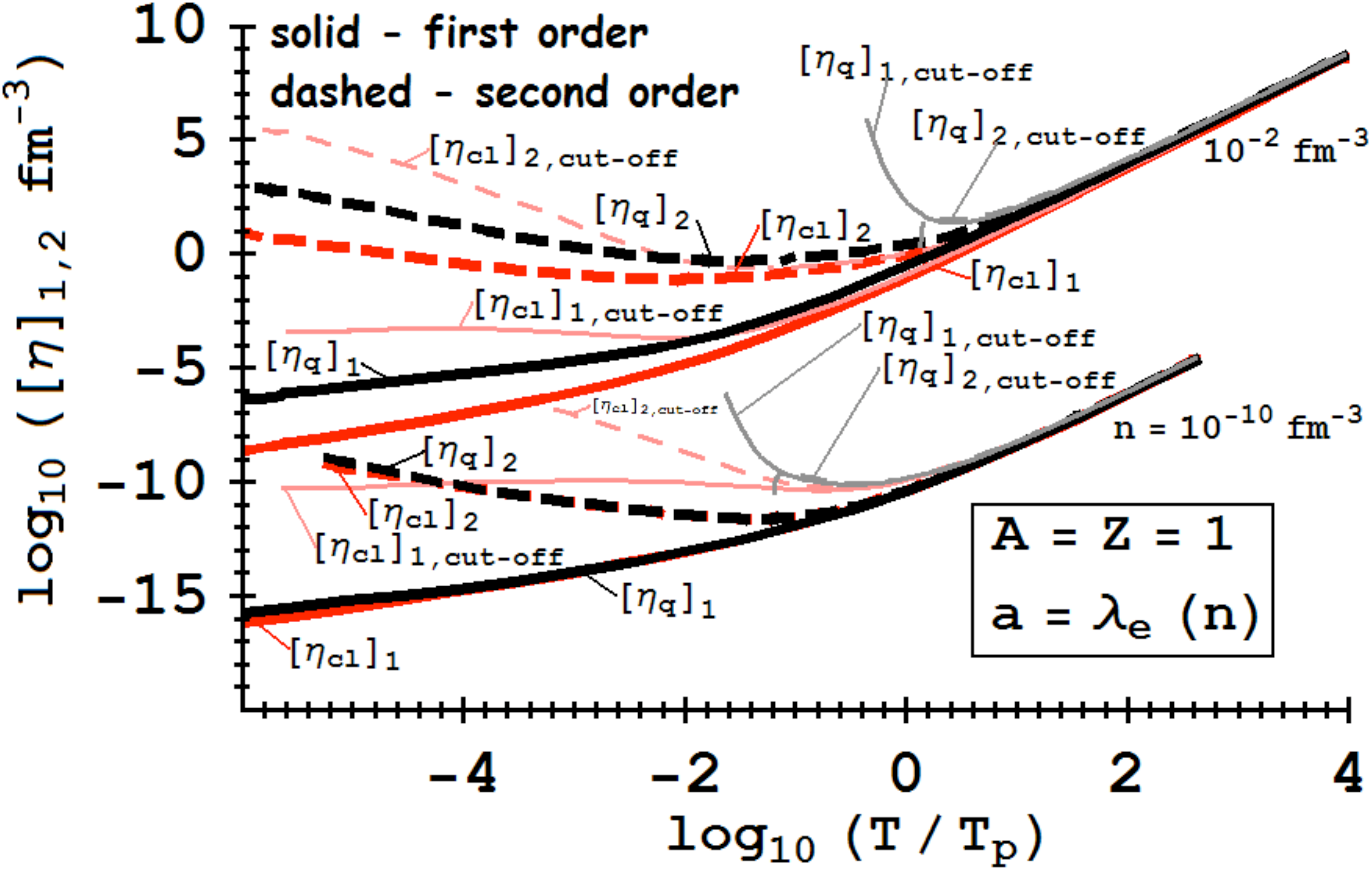}\\
\caption{(Color online.) Viscosities for density dependent screening
  lengths as functions of $T/T_p$ for the Hydrogen plasma (spin = $1/2$).}
	\label{two}
\end{figure}
\end{center}

Additional insight can be gained by examining viscosities as functions of
$T/T_p$ (see Fig.~\ref{two}). Comparing the results for the two
densities shown in this figure, we learn that the quantum and
classical results begin to differ from each other at progressively
lower values of $T/T_p$ as the density decreases.  Whereas the first order
classical and quantum results differ by small amounts, the differences
grow with increasing density. Many body effects gauged through
cut-offs appear together with second order corrections.

An important feature that emerges from the above results is that the
extent to which the quantum results differ from the classical results
is not as large as in the case when $a$ was fixed for all densities as
in Secs. \ref{dilute} and \ref{quantum}. The feedback offered by the
density dependence of the screening length (this is in fact a
many-body effect) serves to reduce the differences between the quantum
and classical viscosities.

In general, quantum effects are important for light ions and for small
screening lengths $\lambda_e$.  In the limit $\lambda_e \rightarrow
\infty$ the potential reduces to a $1/r$ Coulomb potential for which
the classical and quantum differential cross sections agree.  For
Hydrogen with $Z=1$, the screening length from Eq.~(\ref{lambdae}) can
be much larger than the fixed $\lambda_e=10$~ fm used in the previous
sections.  This larger screening length leads to smaller quantum
effects.  (Note that the MD simulations are easier for $\lambda_e=10$
fm.)  We conclude that quantum effects are probably small for
realistic screening lengths.
                                                                               
%\vspace*{3cm}
%                                       

%\vspace*{3cm}
\begin{center}
\begin{figure}[tb]
	\includegraphics[width=3in]{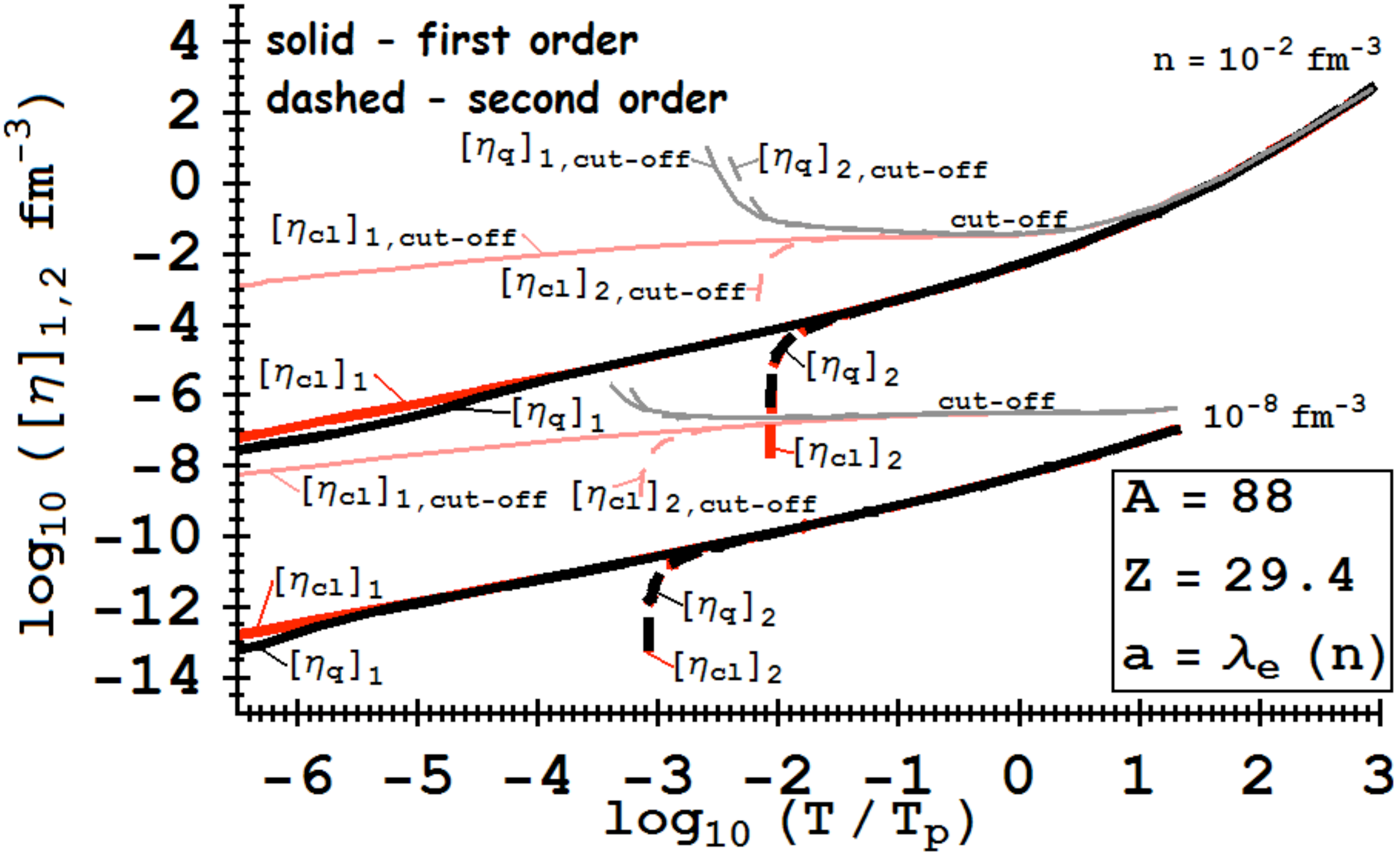}\\
\caption{(Color online.) Same as Fig.~\ref{two}, but for the system
  with $A=88$ (spin = $0$).}
	\label{tri}
\end{figure}
\end{center}

Figure \ref{tri} shows results of viscosities as functions of $T/T_p$
for the system with $A=88$ at two different densities. Whereas the
first order 
classical and quantum results agree over a wide range of $T/T_p$, the
role of many-body effects studied through the influence of
cut-offs is evident even for $T/T_p \gg 1$. Effects of second-order
corrections to viscosity enter at much lower $T/T_p$'s than the effects
of cut-offs. Note that the effects of statistics are opposite for spin
$1/2$ (Fig.~\ref{two}) and spin $0$ systems. 

\section{Comparison with the electron contribution}
\label{electron-shear}
The shear viscosity of the outer crust of a neutron star is determined
by the contribution of the various matter components. In this way the
total shear viscosity $\eta_{tot}$ is given by the sum of the
viscosity due to electrons $\eta_e$ and the ions
$\eta$, $\eta_{tot}=\eta_e +\eta$. In this section we calculate the
electron contribution for a single case, and compare with the ion
contribution obtained in the previous section.  We follow the
formalism used by Chugunov and Yakovlev \cite{Chugunov-shear}. The
shear viscosity coming from the electrons can be calculated from

\begin{equation}
\eta_e=\frac{n_ek_Fv_F}{5\nu_e},
\end{equation}
where $\nu_e$ is the effective electron collision frequency and $v_F$
($\approx 1$) is the electron Fermi velocity.  For dense matter this
collision frequency is given by
\begin{equation}
\nu_e=\nu_{ei}+\nu_{imp}+\nu_{ee},
\end{equation}
where $\nu_{ei}$, $\nu_{imp}$, and $\nu_{ee}$ correspond to electron
scattering from ions, impurities, and electrons, respectively.
We assume electron-ion scattering is the dominant process. The 
electron-ion collision  frequency is given by 

\begin{equation}
\nu_{ei}=\frac{4Z\epsilon_F \alpha^2}{\pi\hbar}\Lambda_{ei},
\end{equation}
 where $\Lambda_{ei}$ is the effective Coulomb logarithm,
 \begin{equation}
 \Lambda_{ei}=\int\limits^{2k_F}_{q_0}
 q^3\frac{u^2(q)}{\epsilon(q,0)^2} 
\left ( 1-\frac{q^2}{4k^2_F}\right)
\left [1-\frac{1}{4}\left(\frac{q}{m_e^*}\right)^2 \right] S_\eta(q)dq,
 \end{equation}
where $q$ is the momentum trasnfer. The lower limit $q_0$ is $0$ in a
liquid phase and $q_0=(6\pi^2n)^{1/3}$ in a crystal phase
\cite{Potekhin-thermal}, the effective electron mass is
$m_e^*=\epsilon_F=(k_F^2+m_e^2)^{1/2}$ with $k_F$ the electron Fermi
momentum and $m_e$ the electron mass. $S_\eta(q)$ is the structure
factor describing electron-ion scattering, $u(q)$ is the Coulomb
interaction between an electron and a nucleus, and $\epsilon(q,0)$ is
the dielectric function due to degenerate relativistic electrons,
\cite{Jancovici-dielectric, itoh-epsilon}.

We calculate $S_\eta(q)$ directly as a density-density correlation
function using trajectories from our MD simulations,
\begin{equation}
S_\eta({\bf q})=\langle \rho^*({\bf q})\rho({\bf q}) \rangle - 
|\langle \rho({\bf q})\rangle |^2\, .
\label{S(q)}
\end{equation}
Here the ion density $\rho({\bf q})$ is,
\begin{equation}
\rho({\bf q}) = \frac{1}{\sqrt N} \sum_{i=1}^N 
 {\rm e}^{i {\bf q \cdot r}_i}.
\label{rho}
\end{equation}

\begin{figure}
\begin{center}
\includegraphics[width=3.5in]  {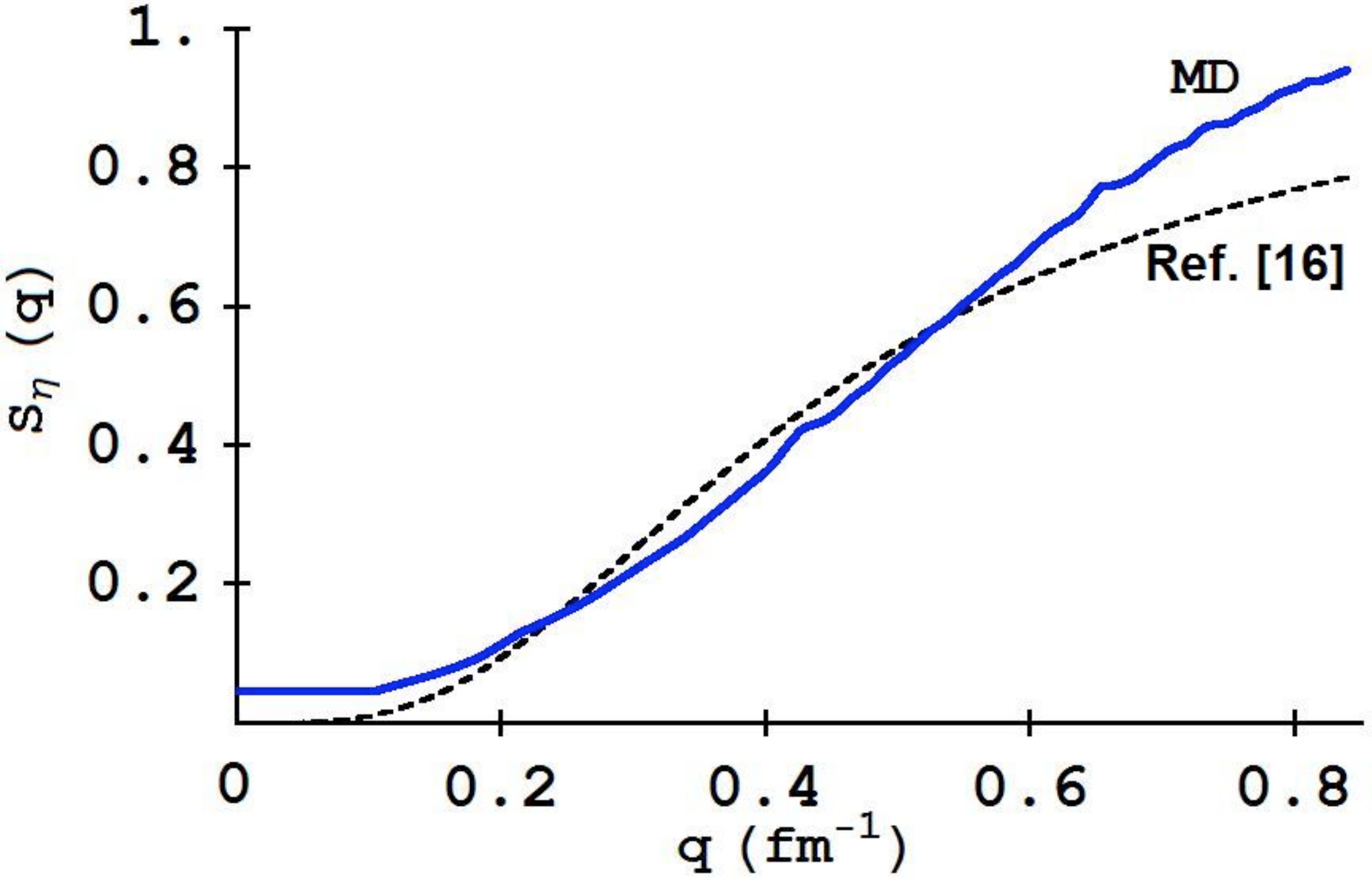}
\caption{(Color online.) Structure factor from MD simulations. The
temperature is 0.1 MeV and $n=2.5\times 10^{-3}$ fm$^{-3}$.  The
dotted curve is an anlytical fit from Ref. \cite{Potekhin-thermal}.}
\label{sqinterp}
\end{center}
\end{figure}

Ref. \cite{Chugunov-shear} calculates the electron contribution to the
shear viscosity using the analytical fit for the effective structure
factor proposed in Ref. \cite{Potekhin-thermal} plus some corrections
included to take into account the form factor of the nucleus.  As
shown in \cite{thermalcond} our MD results for the Coulomb logarithms
are comparable to the analytical results of
Ref. \cite{Potekhin-thermal}.  The structure factor used to find the
Coulomb logarithms is shown in Fig. \ref{sqinterp} along with the
anlytical fit from Ref. \cite{Potekhin-thermal}. The lowest value of
$q$ at which we can calculate $S_\eta(q)$ is limited by
$q_0=2\pi/L$. We approximate $S_\eta(q)$ by $S_\eta(q_0)$ for the low
$q$ region from $0$ to $q_0$. This value agrees with the
Debye-H\"{u}ckel approximation at $q=0$.
  
Table \ref{vis_comp} shows the viscosity  obtained for a
simulation with 500 ions at a temperature of 0.1 MeV and an ion
density of $2.5 \times 10^{-3}$ fm$^{-3}$. 
Even with large quantum corrections, 
the contribution from the
ion-ion scattering is small by orders of magnitude compared with the
contribution from electron-ion scattering. Therefore, for this
regime of densities and temperatures the ion contribution to the shear
viscosity of the outer crust in a neutron star is
negligible. Nevertheless, the use of MD simulations has allowed us to
calculate both contributions.

Our MD formalism can also be used to calculate the thermal
conductivity $\kappa$ \cite{thermalcond}. Again we expect the electron
contribution to $\kappa$ to dominate over the ion contribution at low
magnetic fields. However, large magnetic fields suppress the electron
contribution in directions perpendicular to the field; then the ion
contribution can be important. For example, Potekhin and Yakovlev
\cite{Potekhin01} find that magnetic fields larger than $10^{12}$
Gauss can lead to anisotropic temperature distributions.  Therefore we
expect a field of $10^{12}$ Gauss to start modifying electron
contributions to the shear viscosity.  However, it may take
significantly stronger fields before the ion contribution to the
viscosity dominates that from the electrons.

\begin{table}
\begin{center}
\caption{Contributions to the shear viscosity $\eta$ (for the Hydrogen
  plasma), coming from electron-ion scattering $\eta_{e}$, and from
  ion-ion correlations $\eta$. The system is at a temperature of
  $0.1$ MeV and the ion density is $2.5\times 10^{-3}$ fm$^{-3}$}.\\
 \begin{tabular}{llllllll}
\hline
$ \Lambda_{ei}$ (fm$^{-3}$ )&$\eta_{e}$(fm$^{-3}$)&$\eta$(fm$^{-3}$) \\
%OCP Fit&0.24&29.81& ----\\
0.27&26.61& $4.81\times 10^{-4}$\\
\hline
\end{tabular} 
\label{vis_comp}
\end{center}
\end{table}

\section{Summary and Conclusions}
\label{conclusions}
We have calculated the ion contribution to the shear viscosity in the
outer crust of neutron stars. We use two different methods for this
purpose. One of these is based on MD simulations while the other is
semi-analytical and calculates the transport cross sections for
classical and quantum systems.  We find good agreement between the two
methods in the low density, classical limit.

Using the MD trajectories, we have used autocorrelation functions of
the pressure tensor in a Hydrogen OCP to calculate its viscosity. We
have studied the case in which the plasma was dilute. In this case the
coupling between the ions is weak. We have found that the viscosity is
constant at low densities and then becomes an increasing function of
density, contrary to the prediction including only binary collisions,
where the viscosity remains constant with density.  This fact is due to
the correlations between ions. The dependence of the plasma viscosity
with temperature, at a fixed density, is the one expected for a
gas. However, as the temperature decreases changes in the viscosity
with density, are larger due to stronger ion correlations.

In the dilute gas limit and for light ion systems, calculations based
on the Chapman-Enskog formalism, indicate that viscosities with
quantum transport cross sections are nearly a factor of ten larger
than the classical results for a fixed screening length of 10 fm.  
However, for larger more realistic screening lengths, the difference
between quantum and classical calculations is smaller.

Using the structure factor of the ions we calculate the contribution
to the shear viscosity due to electron-ion scattering. We find that
the contribution to the shear viscosity from the ions is
negligible compared to the former one for the conditions of interest
in the outer crust of neutron stars. Therefore, we do not expect major
damping of the r-modes from the ions in the crust.

Our method of calculating the electron contribution via the structure
factor of the ions also allows us to use multi-component plasmas, and
the contribution of electron-impurity scattering will be automatically
included. On the other hand, we have used ion autocorrelation
functions of the pressure tensor, and the Kubo formalism to calculate
the viscosity of the ions. This method could be extended to calculate
other mechanical properties of the ions, like the shear modulus, of
great importance in the understanding of starquakes.

\section*{Acknowledgements}
\label{ack}
The research of O.~L.~C. and C.~J.~H.  was supported in part by DOE
DE-FG02-87ER40365 and by Shared Research grants from IBM, Inc. to
Indiana University.  The research of S.~P. and M.~P. was supported by
the Department of Energy under grant DE-FG02-93ER40756.

\section*{Appendix. Calculation of the phase shifts}

Here, we employ an efficient algorithm proposed by
Klozenberg~\cite{Klozenberg} to calculate the phase shifts.  The
radial part of the Shr\"odinger equation is
\begin{equation}
  u''(r)+\left(k^2-\frac{l(l+1)}{r^2}-\frac{2\mu}{\hbar^2}V(r)\right)u_l(r)= 0,
\label{radial_eqn_u}
\end{equation}
where $\mu$ is the reduced mass, $r$ is the separation distance, $l$
is the angular momentum quantum number, $k$ is the wave number such
that energy $E=\frac{\hbar^2 k^2}{2\mu}$ and $V(r)$ is a spherically
symmetric potential.  Two new functions $a_l(r)$ and $S_l(r)$ are
defined through the relation
\begin{equation}
\label{sl_al}
    u_l(r)=a_l(r)\left(j_l(kr)+S_l(r) n_l(kr)\right),
\end{equation}
such that
\begin{equation}
\label{sl_al_cnd}
    j_l(kr)\frac{da_l(r)}{dr}+n_l(kr)\frac{d}{dr}\left(S_l(r) a_l(r)\right)=0,
\end{equation}
where $j_l$ and $n_l$ are the Ricatti-Bessel functions, the solutions
of the free radial equation (\ref{radial_eqn_u}) when $V(r)=0$. 
%Some of these functions are listed below
%
%\begin{equation}
%\label{RicattiBessel}
%\begin{split}
%    &j_0 (t)=\sin(t), \, n_0 (t)=\cos(t),\\
%    &j_1 (t)=\frac{\sin(t)}{t}-\cos(t), \, n_1 (t)=\frac{\cos(t)}{t}+\sin(t),\\
%    &\ldots
%\end{split}
%\end{equation}
%which are generated by the recurrence formula:
%\begin{equation}
%\label{rec_fl}
%    f_{l+1}(t)=\frac{l+1}{t}f_l(t)-\frac{df_l(t)}{dt},
%\end{equation}
%where $f_l=j_l$ or $n_l$.

With the function $S_l$, Eq. (\ref{radial_eqn_u}) becomes
\begin{equation}
\label{eqn_S}
    \frac{dS_l(r)}{dr} + 
\frac{2\mu V(r)}{\hbar^2 k}\left(j_l(kr)+S_l(r)n_l(kr)\right)^2=0.
\end{equation}
Let us introduce the dimensionless variable $t=kr$ and write
Eq. (\ref{eqn_S}) as \emph{a phase equation}
\begin{equation}
\label{eqn_S_t}
    \frac{dS_l(t)}{dt}+v_k(t)\left(j_l(t)+S_l(t)n_l(t)\right)^2=0,
\end{equation}
where $v_k(t)=\frac{2\mu V(t/k)}{\hbar^2 k^2}$.  By the construction
in Eq. (\ref{sl_al}), the tangent of the phase shift is the limiting
value of $S_l(t):$
\begin{equation}
\label{Sl_to_tgdl}
    \lim_{t\to\infty}S_l(t)=\tan(\delta_l(k)),
\end{equation}
where the existence of the limit is assured by the fact that the
potential $V(r)$ falls sufficiently rapidly as $r\to \infty$.  When
the phase shift is close to $\frac{\pi}{2}$ plus multiples of $\pi$,
we consider the function $\tilde{S}_l(t)=1/S_l(t)$ and its
differential equation
\begin{equation}
\label{eqn_Stilde_t}
    \frac{d\tilde{S}_l(t)}{dt}-v_k(t)
\left(\tilde{S}_l(t) j_l(t)+n_l(t)\right)^2=0
\end{equation}
to avoid large values of $S_l$ and its infinities.  To treat the
singularities associated with the Ricatti-Bessel functions 
$n_l(t)$ when $t\to0$, the normalized analogues
\begin{equation}
\label{RicattiBessel_norm}
\begin{split}
    &\hat{j}_l(t)\equiv\frac{j_l(t)}{t^{l+1}}, \, 
\hat{n}_l(t)\equiv n_l(t) t^{l},\\
    &\hat{S}_l(t)\equiv\frac{S_l(t)}{t^{2l+1}}, \, 
\hat{\tilde{S}}_l(t)\equiv \tilde{S}_l(t) t^{2l+1},
\end{split}
\end{equation}
are used with the corresponding equations
\begin{equation}
\label{eqn_Shat_t}
    \frac{d\hat{S}_l(t)}{dt}+\frac{2l+1}{t}\hat{S}_l(t)+t \, 
v_k(t)\left(\hat{j}_l(t)+\hat{S}_l(t)\hat{n}_l(t)\right)^2=0,
\end{equation}
\begin{equation}
\label{eqn_Shattilde_t}
    \frac{d\hat{\tilde{S}}_l(t)}{dt}- 
\frac{2l+1}{t}\hat{\tilde{S}}_l(t)-t \, v_k(t)
\left(\hat{\tilde{S}}_l(t)\hat{j}_l(t)+\hat{n}_l(t)\right)^2=0.
\end{equation}
To find the initial conditions, let us expand the solution of
Eq. (\ref{eqn_Shat_t}) around $t=0$ as
\begin{equation}
\label{s_expn}
    \hat{S}_l(t)=c_0+c_1 t+c_2 t^2+O(t^3),
\end{equation}
and assume that the expansion for $v(t)$ is known and has the form
\begin{equation}
\label{v_expn}
    v_k(t)=\frac{a_{-1}}{t}+a_0+a_1 t+O(t^2).
\end{equation}
This leads to the coefficients
\begin{equation}
%\label{c0}
    c_0=0\,, \qquad
%\end{equation}
%\begin{equation}
\label{c1}
    c_1=-\frac{a_{-1}/2}{(l+1)\left((2l+1)!!\right)^2} \, ,
\end{equation}
\begin{equation}
\label{c2}
    c_2=\frac{{a_{-1}}^2}{(2l+3)(2l+1)(l+1)\left((2l+1)!!\right)^2} - 
\frac{a_0}{(2l+3)\left((2l+1)!!\right)} \, ,
\end{equation}
whence
\begin{equation}
\label{init}
    \hat{S}_l(t\ll\left|\frac{c_1}{c_2}\right|)\simeq c_1 t+c_2 t^2,
\end{equation}
which is the initial condition for Eq. (\ref{eqn_Shat_t}). During
numerical integration, we switch to Eq. (\ref{eqn_Shattilde_t}) when
$|\hat{S}_l(t)|>1$, and if $|\hat{\tilde{S}}_l(t)|>1$ we return back
to Eq. (\ref{eqn_Shat_t}).  This procedure is carried out until $t$
reaches unity, when we switch to the set of Eqs. (\ref{eqn_S_t}) and
(\ref{eqn_Stilde_t}).  After the solution reaches its limiting value
of $\tan(\delta_l(k))$ in Eq.~(\ref{Sl_to_tgdl}) to satisfactory
precision, the evaluation is stopped.

\subsection*{JWKB phase shifts}

The JWKB approximation allows fast and simple
calculations of phase shifts for large $k-$ and moderately large
$l$-values.  
Here we briefly describe this method following Mott and
Massey~\cite{Mott}, 
and, Cohen~\cite{Cohen}.  Let us write the wave equation as
\begin{equation}
  u_l''(r)+F(r) \, u_l(r)= 0,
\label{wave_eqn_u}
\end{equation}
\begin{equation}
  F(r)=\frac{2\mu}{\hbar^2}(E-V(r))-\frac{l(l+1)}{r^2}.
\label{Fr}
\end{equation}
If the particle's energy and angular momentum are sufficiently large,
such that the potential energy varies only slightly over a few 
wavelengths, then in the limit  $\hbar \to 0$ we may suppose that
$F$ is large. Then the solutions of Eq. (\ref{wave_eqn_u}) are
approximately
\begin{equation}
u_l\approx F^{-\frac{1}{4}} \exp\left[\pm i \int^r F^{\frac{1}{2}} dr \right].
\label{Fsol}	
\end{equation}
Jeffreys~\cite{Jeffreys} has shown that the solution which decreases
exponentially inside the classically accessible region ($r<r_0$) is
\begin{equation}
u_l(r)\approx F^{-\frac{1}{4}} \sin
\left[\frac{\pi}{4}+\int_{r_0}^r F^{\frac{1}{2}} dr \right],
\label{jef}	
\end{equation}
where $r_0$ is the classical turning point given by the positive root
of $F(r_0)=0$.  To find the solution which vanishes as $r\to 0$,
Langer~\cite{Langer} employed the substitutions $\rho=\log(r)$ and
$g_l=\frac{u_l(r)}{\sqrt{r}}$ in Eq. ({\ref{wave_eqn_u}}) and obtained
\begin{equation}
 \frac{d^2 g_l}{d\rho^2}+e^{2\rho} F_1(r) g_l= 0,
\label{langer_eq}
\end{equation}
\begin{equation}
  F_1=\frac{2\mu}{\hbar^2}(E-V)-\frac{(l+\frac{1}{2})^2}{r^2},
\label{F1}
\end{equation}
and by Jeffrey's method, the solution is
\begin{equation}
u_l(r)\approx F_1^{-\frac{1}{4}} \sin
\left[\frac{\pi}{4}+\int_{r_0}^r F_1^{\frac{1}{2}} dr \right],
\label{jef1}	
\end{equation}
where now $F_1(r_0)=0$.

%\vspace*{3cm}
\begin{center}
\begin{figure}[tb]
	\includegraphics[width=3in]{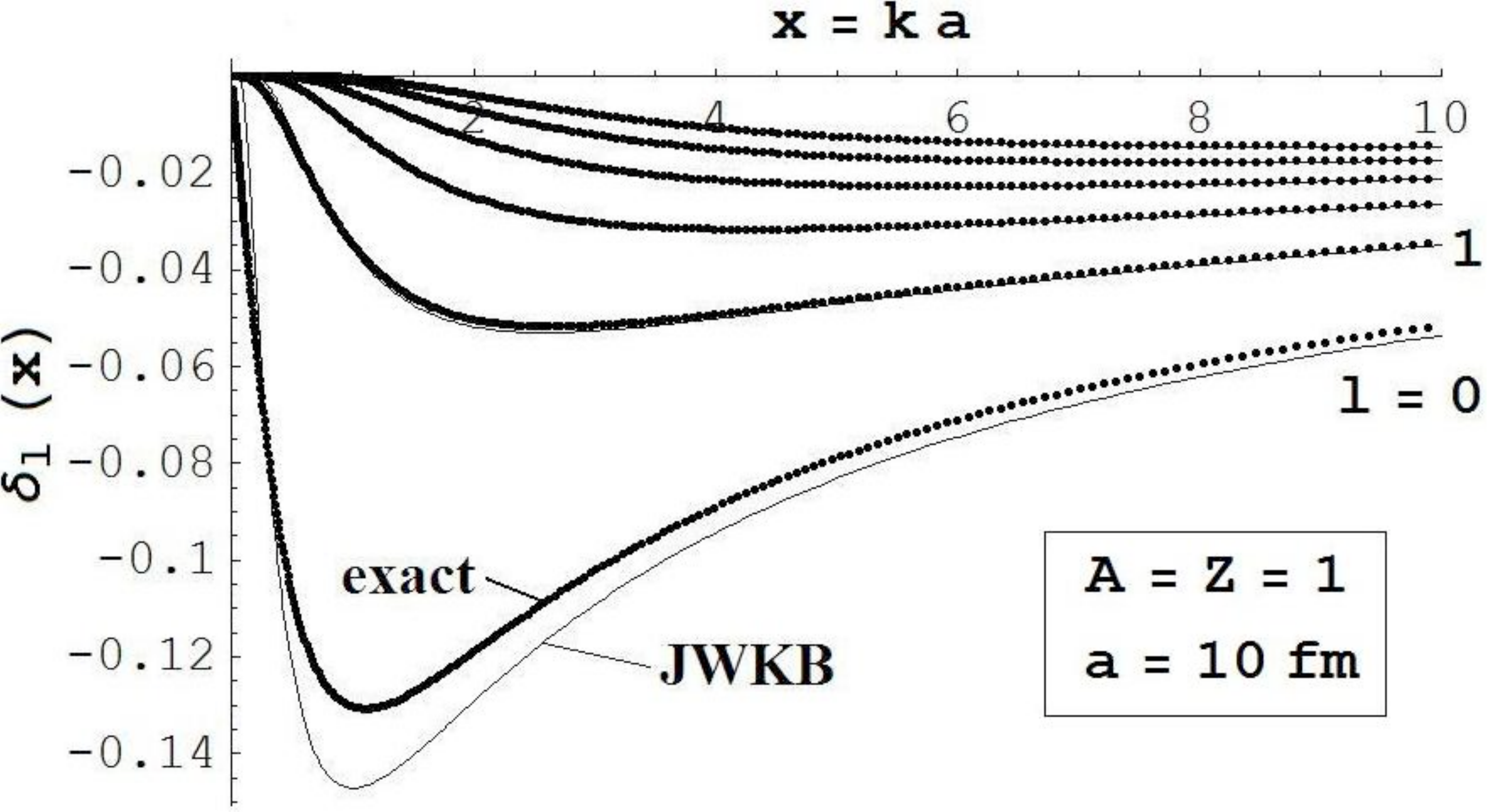}\\
\caption{(Color online.) Exact and JWKB phase shifts for the Yukawa
  potential in Eq.~(\ref{potential}) for a system with $Z=A=1$ and
  $a=10$ fm.}
	\label{ex_JWKB_figure}
\end{figure}
\end{center}
For large $r$, the asymptotic form of Eq. (\ref{jef1}) is
\begin{equation}
 \sin\left[ \frac{\pi}{4}+\int_{r_0}^{\infty} 
(F_1^{\frac{1}{2}}-k)dr + k(r-r_0) \right] = 
\sin\left[kr-\frac{l\pi}{2}+\delta_l\right],
\label{asym}	
\end{equation}
whence
\begin{equation}
\delta_l = \frac{\pi}{4} + \frac{l\pi}{2}-k r_0 + 
\int_{r_0}^{\infty} (F_1^{\frac{1}{2}}-k)dr.
\label{jwkb_shifts}
\end{equation}
Finally, following Cohen~\cite{Cohen}, we rewrite the above equation in the
form used for numerical evaluations:
\begin{eqnarray}
\delta(b,k) &=& k\int_{r_0(b)}^{\infty}
\left[\sqrt{1-\frac{V(r)}{E} - \frac{b^2}{r^2}} - 
\sqrt{1-\frac{r_0^2(b)}{r^2}}\right]dr \nonumber \\ 
&+& \frac{k(b-r_0(b))\pi}{2},
\label{jwkb-frml}
\end{eqnarray}
where $b=\frac{l+1/2}{k}$ is the classical impact parameter.

For numerical computations, we choose two sets of parameters for the
Yukawa potential in Eq.~(\ref{potential}): $Z=A=1$ and $a=10$ fm, and,
$Z=29.4$, $A=88$ and $a=26.1$ fm.  An exact calculation of the phase
shifts is performed for $l=0-5$ in the range $x=0-1000$. For larger
values of $l$, up to $l=100$ for $A=1$ and up to $l=500$ for $A=88$,
the phase shifts are computed using the the JWKB approximation.  As
shown in Fig. \ref{ex_JWKB_figure}, the JWKB results are very close to
the exact values, even for $l=1$. The agreement is better for the
system with $A=88$ (not shown here), which is essentially classical.
%

%%%%%%%%%%%%%%%%%%%%%%%%%%%%%%%%%%%%%%%%%%%%%%%%%%%%%%%%%%%%%%%%%
\vfill\eject
\end{document}